\documentclass[11pt,final]{article}

\usepackage{amssymb}
\usepackage{amsmath}
\usepackage{graphicx}
\usepackage[margin=1.0in]{geometry}

\title{Stochastic subgrid-scale parameterization for one-dimensional shallow water dynamics using stochastic mode reduction
\footnote{Submitted to Q.J.R. Meteorol. Soc.}}

\date{\today}

\author{Matthias Zacharuk
\thanks{Corresponding author; Institut f\"ur Atmosph\"are und Umwelt, Fachbereich 
Geowissenschaften/Geographie, Goethe-Universit\"at (zacharuk@iau.uni-frankfurt.de).}
\and
Stamen I. Dolaptchiev
\thanks{Institut f\"ur Atmosph\"are und Umwelt, Fachbereich 
Geowissenschaften/Geographie, Goethe-Universit\"at.}
\and
Ulrich Achatz
\footnotemark[3]
\and
Ilya Timofeyev 
\thanks{Department of Mathematics,
University of Houston,
Houston, TX 77204-3008, (\texttt{ilya@math.uh.edu}).}
}

\begin{document}

\maketitle

\begin{abstract}
We address the question of parameterizing the
    subgrid scales in simulations of geophysical flows by applying
    stochastic mode reduction to the one-dimensional stochastically
    forced shallow water equations. The problem is formulated in
    physical space by defining resolved variables as local spatial
    averages over finite-volume cells and unresolved variables as
    corresponding residuals. Based on the assumption of a time-scale
    separation between the slow spatial averages and the fast
    residuals, the stochastic mode reduction procedure is used to
    obtain a low-resolution model for the spatial averages alone with
    local stochastic subgrid-scale parameterization coupling each
    resolved variable only to a few neighboring cells. The closure
    improves the results of the low-resolution model and outperforms
    two purely empirical stochastic parameterizations. It is shown
    that the largest benefit is in the representation of the energy
    spectrum. By adjusting only a single coefficient (the strength of
    the noise) we observe that there is a potential for improving the
    performance of the parameterization, if additional tuning of the
    coefficients is performed. In addition, the scale-awareness of the
    parameterizations is studied.
  %

\end{abstract}

%
\section{Introduction} 
Atmospheric processes encompass a large spectrum of spatial and
temporal scales.  These range from several millimeters and seconds for
boundary layer turbulence up to \(10^7\) meters and several weeks (and
even longer) for planetary wave dynamics. Due to limited computer
resources numerical atmospheric models cannot describe all these
processes on all scales simultaneously.  
However, the different scales are interacting in a complex manner and
this leads to the challenging problem of parameterizing the effect of
the unresolved subgrid-scale (SGS) processes onto the resolved ones.
Examples include the parameterization of synoptic and mesoscale eddies
in planetary scale atmospheric models (e.g. \cite{petoukhov_00,Weaver2001}), 
momentum and temperature fluxes in the atmospheric
boundary layer \cite{Stull1988} or SGS Reynold stresses in large eddy
simulations (e.g. \cite{Pope2000}).

In this context, stochastic elements have become increasingly popular.
Stochastic parameterizations can reduce a systematic model error,
represent uncertainty in predictions, or trigger regime transitions
(e.g. \cite{Palmer2001, Berner2017}). Typically some {\sl ad-hoc}
SGS model is assumed and the corresponding coefficients are optimized
(tuned) so as to obtain the best possible agreement, in some sense,
with observations or high-resolution simulations.  Examples in
comprehensive climate and weather models are stochastically perturbed
parameterization tendencies \cite{Buizza1999, palmer2009} or
stochastic kinetic energy backscatter \cite{Shutts2005, Berner2009}.
Empirical Ornstein-Uhlenbeck (OU) processes have been used in some
studies of low-frequency and large-scale atmospheric variability
(e.g. \cite{Winkler2001, Newman2003, Pegion2011}), which can be
extended to include quadratic nonlinearities as well as a time
correlated stochastic forcing \cite{Kravtsov2005, Kondrashov2005}.

With regard to SGS parameterizations in climate models issues can
arise from the fact that they are typically tuned to optimally
represent the statistics of the present-day climate. If climate
changes due to some external forcing, it is not guaranteed that the
tuned parameters are still optimal. The fluctuation-dissipation
theorem might be able to provide corrections \cite{Achatz2013,pieroth} 
in some cases, but such an approach relies on the
perturbations being sufficiently weak. Moreover, there is a need for
scale-aware parameterizations in atmosphere modeling, as model
resolution increases continuously and mesh refinement techniques
become widely used. In addition, the consistency between particular
SGS parameterizations and the numerical discretization becomes
important.

These considerations motivate the development of other approaches
where the SGS parameterization is derived from first principles, if
possible without any empirical parameter optimization.  The
  direct interaction approximation (DIA) introduced by \cite{Kraichnan_59}
  allowed to successfully apply statistical dynamical closure theory
  in relevant geophysical flows \cite{frederiksen_97,frederiksen_99,frederiksen_03}.
  In the presence of time scale separation, the
asymptotic method of averaging has been applied \cite{Hasselmann1976,
  Imkeller2001, Arnold2003, Monahan2011}.  This method requires an
estimation of the invariant measure of the fast scales conditioned on
the slow scales, which might limit its applicability when going to
high-dimensional systems. Another promising approach, without any
empirical component, is based on the maximum entropy principle
\cite{Verkley2011, Verkley2014,Verkley2016}. Recently,
\cite{wouters2012, wouters2013} have introduced a new method
originating from response theory. This method relies on a weak
coupling between resolved and unresolved scales and it has been
applied to simple and more complex settings
\cite{wouters2016,Demaeyer2017b,Vissio2017}.

The DIA parameterization has been successfully applied to
  barotropic \cite{frederiksen_97} and primitive equations model
  \cite{frederiksen_03}, it has been extended to include the effects
  of mean flow and topography \cite{frederiksen_99}. The DIA closure
  is derived in spectral space by considering the evolution of
  second-order cumulant and response function. Next, the nonlinear
  damping rate and nonlinear noise are introduced. This results in a
  globally coupled SGS model in spectral space. However, techniques
  have been proposed to simplify the equations and obtain locally
  coupled models reproducing the spectra from direct numerical
  simulations \cite{frederiksen_97, frederiksen_03}.

Another nearly self-consistent possibility that exploits a separation
of time scales between resolved and unresolved scales is the
stochastic mode reduction (SMR) procedure proposed by \cite{Majda2001,
  Majda2002, Majda2003a}. The SMR is a homogenization technique for
multiscale systems (\cite{kha66a,kha66b,kur73,pap76} and a recent
overview \cite{Pavliotis2008} and references therein) and it is
supplemented by an empirical step, where the fast SGS
self-interactions in the evolution equation for the unresolved modes
are replaced by an OU process.  Following this step, an analytical
derivation of a stochastic parameterization for the fast modes is
possible, rigorously valid in the limit of infinite time scale
separation.

So far the SMR procedure has already been applied to balanced models,
such as barotropic \cite{Franzke2005} and quasi-geostrophic
\cite{Franzke2006} dynamics. The separation between resolved and
unresolved scales has been performed by using empirical orthogonal
functions (EOFs). However, EOFs are sometimes not able to
guarantee a sufficient separation of the underlying time scales
(e.g. Figure~3 in \cite{Franzke2006}).
The SMR carried out in spectral space
  \cite{Franzke2005, Franzke2006} is quite similar to the DIA closure approach.
  In particular, the main goal of both techniques is to represent the
  subgrid processes by a nonlinear damping and a state-dependent
  noise and both techniques have been utilized successfully in
  geophysical flows.

%
In applications of the SMR to spectral space the resulting
  reduced model is globally coupled with linear, quadratic and even
  cubic terms. This hampers the applicability of the technique when
  high-dimensional systems with large number of resolved modes are
  considered. However, the latter problem can be avoided by applying the
  SMR in physical space to a finite-volume discretization of the
  equations.
Such discretization does not per se include global coupling as
  in spectral discretizations, since grid cells interact directly only
  with a small number of neighbors. Finite-volume schemes are
  traditionally applied in ocean models (e.g. \cite{Haidvogel1999}),
  regional atmospheric modeling (e.g. \cite{wrf}) and
  recently even for global atmospheric models as well
  \cite{Ripodas2009, Majewski2002, Satoh2008}.  In the examples
    above complex boundaries, such as continental boundaries in ocean
    modeling or non-periodic lateral boundaries in regional area
    atmospheric modeling, necessitate the use of discretizations and
    SGS parameterizations formulated in physical space.
  This motivated \cite{Dolaptchiev2013, Dolaptchiev2013a} to consider
  a local approach where the resolved variables are defined by local
  spatial averages and the SGS flow by deviations from these averages,
  a configuration typically encountered in large-eddy turbulence
  parameterization (e.g. \cite{Pope2000}). The local definition
  leads to a local SGS parameterization, coupling only near neighbors,
  as shown for the Burgers equation \cite{Dolaptchiev2013,
    Dolaptchiev2013a}. The efficient local stochastic SGS
  parameterization allows to consider large numbers of resolved
  scales. In addition, the clear gap of spatial scales between the
  resolved and unresolved variables enables a more pronounced
  time-scale separation.

Obviously Burgers equation represents a highly idealized prototype
model for testing various statistical and closure methods and it is
necessary to verify the applicability of the SMR for local spatial
averages for more realistic fluid-dynamical models.  One step in this
direction is performed in this work by applying the approach to a
stochastically forced one-dimensional shallow water layer (1DSW).  It
incorporates at least two issues of relevance in the general context.
First, in contrast to the Burgers equation the 1DSW allows for gravity
waves. Secondly, if formulated in flux form, the shallow-water flow
dynamical equations entail non-polynomial nonlinearities.  This
problem is of broader relevance, since such highly nonlinear terms
appear in the general compressible fluid flow equations as well, in
the pressure-gradient acceleration.

The work presented here can be summarized as follows. Based on a
high-resolution finite-volume discretization of the shallow-water
equations we use in Sec.~\ref{SEC_Method} local spatial averages to
define coarse and slow (resolved) variables and, via corresponding
residuals, fine and fast (unresolved) variables. The assumed
time-scale separation is verified numerically. The SMR theory for
obtaining an SGS parameterization of the unresolved modes is then
introduced and applied to the specific problem. In
Sec.~\ref{SEC_Models} we discuss the practical implementation, and
also introduce, for comparison, two purely empirical approaches.
Results from model simulations with the various SGS parameterizations
are then compared in Sec.~\ref{SEC_Results}. Here we also investigate
the scale awareness of the approaches, i.e. their ability to be
applied at different resolutions without any re-tuning. Conclusions
are finally drawn in Sec.~\ref{SEC_Conclusion}.

\section{Method \label{SEC_Method}}
\subsection{Shallow water model \label{SEC_1DSW}}
We consider a stochastically forced one dimensional shallow water
layer  with periodic boundaries, using as variables the height of the
fluid \(h\) and the momentum \(hu\), where \(u\) is the velocity. The
governing equations with plane topography 
(e.g. \cite{Vallis2006}) read in flux form

\begin{align}
\partial_t \left( \begin{array}{c}
h \\ hu
\end{array} \right)
=- \partial_x  \left(\begin{array}{c}
hu - \nu \partial_x h \\  \frac{(h u)^2}{h} + \frac{g}{2} h^2 -
\nu \partial_x hu  \end{array}\right) + \pmb{\varrho} ,
\label{SW}
\end{align}
with a large-scale stochastic forcing \(\pmb{\varrho}\) (see Sec.
\ref{SEC_Models}) and a mass weighted diffusion with the constant
parameter \( \nu \).

For a high-resolution spatial discretization the domain of length
\(L\) is divided into \(N\) fine intervals \(\Delta x = L/N\) ,
labelled by a small index \(i \in \{0,1,...,N-1\}\).  With this
the equations in \eqref{SW} can be discretized by a symmetric
finite-volume scheme
\begin{align}
\frac{d}{dt} \left( \begin{array}{c} h_i \\ hu_i \end{array} \right)
 = - \frac{1}{\Delta x } \left( \textbf{F}_{i + \frac{1}{2}} - \textbf{F}_{i-\frac{1}{2}} \right) + \pmb{\varrho}_i  \; ,
\label{DGL dicretised}
\end{align}
with the discrete forcing \(\pmb{\varrho}_i\) and the flux at the boundary given by
\begin{align}
\label{flux_i}
\textbf{F}_{i+\frac{1}{2}}=
\frac{1}{2}
\left(  \begin{array}{c}
\displaystyle{(hu)_{i+1} + (hu)_{i} - 2\nu \frac{h_{i+1} - h_{i}}{\Delta x}}
\\
\displaystyle{\frac{(hu)_{i+1}^2}{h_{i+1}} + \frac{(hu)_{i}^2}{h_{i}} + \frac{g}{2} h_{i+1} ^2 + \frac{g}{2} h_{i} ^2 - 2\nu \frac{(hu)_{i+1} - (hu)_i}{\Delta x}}
\end{array} \right)  \;.
\end{align}

The discrete flux form \eqref{DGL dicretised} conserves total mass \(
\frac{1}{N} \sum_{k=0} ^{N-1} h_k\) and total momentum \( \frac{1}{N}
\sum_{k=0} ^{N-1} hu_k\) in the absence of forcing.  Given our choice
of forcing and dissipation, the number of fine cells is chosen large
enough so as to resolve all processes occurring. Hence in the
following simulations using \eqref{DGL dicretised} , with $N$ large
enough, will be called direct numerical simulation (DNS).

\subsection{Local averages  \label{SEC_Local-averages}}
As a representation of the typical situation of atmospheric models
with insufficient resolution, we introduce a second discretization,
with \(N_c = N/n\) coarse cells, each consisting of \(n\) fine cells
of the initial discretization, and labelled by the capital index \(I
\in \{0,1,...,N_c-1\}\). Associated with the coarse grid coarse
variables $H$ and $HU$, also called resolved variables, are defined by local
spatial averages inside a coarse box

 \begin{align}
\left( \begin{array}{c} H_I \\ HU_I \end{array} \right) &=
 \frac{1}{n} \sum_{k=nI} ^{n(I+1)-1} \left( \begin{array}{c} h_k \\ hu_k \end{array} \right) \, .
\end{align}

Further, fine variables $h'$ and $hu'$, referred to as
  unresolved or SGS variables, are defined using the deviations of the
  initial variables from the corresponding coarse variables


 \begin{align}
 \left( \begin{array}{c} h'_i \\ hu'_i \end{array} \right) 
= \left( \begin{array}{c} h_i \\ hu_i \end{array} \right) 
- \left( \begin{array}{c} H_{I[i]} \\ HU_{I[i]} \end{array} \right)\, .
\end{align}
Here
\(I[i]\) denotes here the index of the coarse cell with the \(i\)-th fine cell placed inside.
The coarse and fine variables can be used to express \eqref{DGL dicretised} as
%
%
\begin{align}
\frac{d}{dt} \left( \begin{array}{c} H_{I} \\ HU_{I} \end{array} \right) 
= & - \frac{\textbf{F}_{(I+1)n - \frac{1}{2} } - \textbf{F}_{nI-\frac{1}{2} }} {n \Delta x} + \pmb{\varrho}_{I} \;,
\label{Split-DGL_x}\\
\frac{d}{dt}\left( \begin{array}{c} h'_i \\ hu'_i \end{array} \right) 
 =& - \frac{\textbf{F}_{i + \frac{1}{2}} - \textbf{F}_{i-\frac{1}{2}}}{\Delta x } + \frac{\textbf{F}_{(I[i]+1)n - \frac{1}{2} } - \textbf{F}_{I[i]n - \frac{1}{2}}} {n \Delta x}  \; ,
\label{Split-DGL_z}
\end{align} 
where we assume that the forcing \(\pmb{\varrho}_I\) acts only onto
the coarse variables. By collecting all resolved variables $H_I, HU_I$
in one vector $\textbf{x} \in \mathbb R^{M_c}
$ and all SGS variables
$h'_i, hu'_i$ in another vector $\textbf{z} \in \mathbb R^{M}$ with
$M_c=2N_c$ and $M=2N$, \eqref{Split-DGL_x} and \eqref{Split-DGL_z}
can be rewritten as
\begin{align}
\dot{x}_i =&  \varrho^x_i +  a^x_i(\textbf{x}) +  b^{xz}_i(\textbf{x},\textbf{z}) \;,
\label{Full eq_x}\\
\dot{z}_i =& b^z_i(\textbf{x},\textbf{z}) + c^z_i(\textbf{z}) \;.
\label{Full eq_z}
\end{align}
Here $\varrho^x_i$ results from the forcing and terms have been
regrouped so as to identify the coarse variable self-interactions
\(a^x_i(\textbf{x})\), the coupling terms
\(b^{xz}_i(\textbf{x},\textbf{z})\),
\(b^{z}_i(\textbf{x},\textbf{z})\) and the fine variable
self-interactions \(c^z_i (\textbf{z})\). Complete neglect
of the SGS variables yields the bare-truncation model
\begin{align}
\dot{x}_i =  \varrho^x_i +  a^x_i(\textbf{x}) \;. \label{Bare-Truncation}
\end{align}
This low resolution model is defined on the coarse grid with \(N_c\)
grid cells and it lacks an SGS parameterization.

\subsection{Stochastic Mode Reduction}
\label{SEC_model-reduction}

\subsubsection{Quadratic approximation \label{SEC_Quadratic-approx}}

A difficulty in the application of SMR is caused by the terms
involving \(1/h\) in the flux \eqref{flux_i} , as they represent a
nonlinearity of an arbitrary order.  So far SMR has only been applied
to systems with quadratic nonlinearities. It can, however, handle
nonlinearities of arbitrary polynomial form. Hence a solution could be
expanding everywhere but in $a^x_I(\textbf{x})$ terms with \(1/h\) in
a finite Taylor series around the mean fluid height \(\mathcal{H}\).
It turns out sufficient, however, to simply replace \(1/h \approx
1/\mathcal{H}\) in order to reproduce the statistics of the DNS.
Thus, the bare truncation part of the model is computed exactly but in
all other terms involving SGS modes this approximation is used,
leading to an approximation of \eqref{Full eq_x} and \eqref{Full
  eq_z}, where SGS-variable nonlinearities take a quadratic form
\begin{align}
\dot{x}_i =&  \varrho^x_i +  a^x_i(\textbf{x}) +
\left( L_{ij}^{xz} z_j + B_{ijk} ^{xxz} x_j z_k +  B_{ijk}^{xzz} z_j z_k \right) ,
\label{Full-h_x}\\
\dot{z}_i =&  \left( L_{ij}^{zx} x_j + B_{ijk}^{zxx} x_j x_k +
B_{ijk} ^{zxz} x_j z_k \right) +
 \left( L_{ij}^{zz} z_j + B_{ijk}^{zzz} z_j z_k \right) .
\label{Full-h_z}
\end{align}
Here and in the following we make use of Einstein's summation
convention and the summation index is running up to either $M_c$ or
$M$ depending on if $\textbf{x}$ or $\textbf{z}$ is involved. The
linear and quadratic interaction coefficients \(L_{ij}^{xz}\),
\(L_{ij}^{zx}\), \(L_{ij}^{zz}\), \(B_{ijk} ^{xxz}\),
\(B_{ijk}^{xzz}\), \(B_{ijk}^{zxx}\), \(B_{ijk} ^{zxz}\),
\(B_{ijk}^{zzz}\) are given in Appendix A.

\subsubsection{Empirical OU process \label{SEC_OU}}

The SGS variables \(z_i\) are not independent,
since the corresponding local spatial average over a coarse cell
vanishes by definition.  Thus, one degree of freedom for each coarse
cell has to be eliminated. This is
  achieved by Fourier-transforming \(h'\) and \(hu'\) locally inside
  each coarse cell. The Fourier amplitude of the zero-wavenumber is
  equal to the vanishing local average inside the coarse cell. Hence
  by discarding this wavenumber component one degree of freedom can be
  eliminated. This defines the new independent SGS variables
$\theta_i$
 \begin{align}
\theta_i &= \hat{T}_{ij} z_j\\
 z_i &= \hat{R}_{ij} \theta_j ,
\end{align}
where the matrices \(\hat{R} \in \mathbb{R}^{M\times M_f} \) and $
\hat{T} \in \mathbb{R}^{M_f\times M} $ are constructed from the
inverse and forward Fourier transformation and $M_f=M-M_c$ is the
number of independent SGS variables. With this preparation one can
move to the next step of SMR, i.e.  replacing the SGS
self-interactions $ L_{ij}^{zz} z_j + B_{ijk}^{zzz} z_j z_k $ by an
empirical OU process. The SGS equation \eqref{Full-h_z} becomes
\begin{align}
d\theta_i =& \hat T_{ik} \left( L_{kj}^{zx} x_j + B_{kjl}^{zxx} x_j x_l +
B_{kjl} ^{zxz} x_j \hat R_{lm}\theta_m \right) dt + \Gamma _{ij} \theta_j dt + \sigma_i dW_i\, , \label{DGL_fine_fourier}
\end{align}
where $\Gamma_{ij}$ denote the coefficients of the negative-definite
OU drift, $\sigma_{ij} = \sigma_i \delta_{ij}$ those of a diagonal
diffusion tensor, and \(dW_i\) Wiener increments. Note that no sum
over $i$ is taken in the Wiener term in (\ref{DGL_fine_fourier}). We
assume that $\Gamma$ couples only SGS modes corresponding to the same
coarse cell. Under this assumption $\Gamma$ has a block-diagonal form
and the resulting SGS closure is local, coupling only neighbors and
next-neighbors of a coarse cell.  Since SMR assumes further that the
OU process is the dominant term in the SGS equation (see below), the
OU drift $\Gamma$ can be estimated from the lagged covariance of $\pmb{\theta}$
\cite{Gardiner2009}
\begin{align}
\overline{\pmb{\theta}(t) \pmb{\theta}^T (t+\tau)} = C(\tau) = C(0) e^{\Gamma^T \tau} ,
\end{align}
where $\overline{\left(\cdot\right)}$ denotes a time average. By integrating over time one can solve for \(\Gamma\)

\begin{align}
\begin{aligned}
 \left(\Gamma^T\right) ^{-1} =& - C(0)^{-1} \int _0 ^{\infty} C(\tau) d\tau  \;.
\end{aligned} \label{OU-parameter_g}
\end{align}

The computation of the time integral in \eqref{OU-parameter_g} is
performed using the numerically efficient Cooper-Haynes algorithm
\cite{Lutsko2015}. Note that despite the block-diagonal form,
$\Gamma$ still allows for a coupling between both SGS variables
\(h'_i\) and \(hu'_i\) inside each coarse cell. Because of the spatial
homogeneity of the considered shallow-water model the coefficients of
$\Gamma$ are the same for each matrix block and can be obtained by
averaging over the estimates from the different coarse cells. Using
the steady Lyapunov equation \( \Gamma C(0) + C(0)\Gamma^T = -\sigma
\sigma^T\), the diagonal diffusion coefficients are found from

\begin{align}
\begin{aligned}
\sigma_i =& \sqrt{ - 2 \Gamma_{ik} C(0)_{ki} } \;.
\end{aligned} \label{OU-parameter_s}
\end{align}

Moreover, it has turned out to be useful to observe that in typical
applications $\Gamma$ is diagonalizable and has distinct eigenvalues
(e.g. \cite{delsole2004}). We hence introduce new variables
\(y_i\)

 \begin{align}
   y_i = U^{-1}_{ij} \theta_j \, ,
\end{align}
where the  real invertible matrix \(U\) is from the real Jordan canonical form decomposition of $\Gamma$

\begin{align}
 \Gamma = U \Lambda U^{-1} \;.
\end{align}

By applying the transformation matrices \(R=\hat R U\) and \(T= U^{-1}
\hat T\), the model equations \eqref{Full-h_x} and 
\eqref{DGL_fine_fourier} can be written in terms of the new variables
as

\begin{align}
\dot{x}_i =& \varrho^x_i + a^x_i (\textbf{x}) +  b^{x}_i(\textbf{x},\textbf{y}) \;,
\label{OU-DNS_x}
\\
\dot{y}_i =& b^y_i(\textbf{x},\textbf{y}) + \Lambda_{ij} y_j + \Sigma_i \dot{W}_i \, .
\label{OU-DNS_y}
\end{align}

Here we use the notation
\begin{align}
  b^{x}_i(\textbf{x},\textbf{y}) =& L_{ij}^{xz} R_{jk} y_k + B_{ijk}
  ^{xxz} R_{kl} x_j y_l + B_{ijk}^{xzz} R_{jl} R_{km} y_l y_m
  \nonumber\\
  =& L_{ij}^{xy} y_j + B_{ijk} ^{xxy} x_j y_k + B_{ijk}^{xyy} y_j y_k\, ,
  \label{eq_b_xi}
  \\
  b^y_i(\textbf{x},\textbf{y}) =& T_{ij} \left( L_{jk}^{zx} x_k +
    B_{jkl}^{zxx} x_k x_l + B_{jkl} ^{zxz} R_{lm} x_k y_m \right)
  \nonumber\\
  =& L_{ij}^{yx} x_j + B_{ijk}^{yxx} x_j x_k + B_{ijk} ^{yxy} x_j y_k
  \; ,
  \label{eq_b_yi}
\end{align}
and we have also introduced effective drift coefficients
\(\Sigma_i=\sqrt{U^{-1} _{ij} U^{-1} _{ij} \sigma_{j}^2 }\) with
pairwise identical noise parameters for pairs of complex eigenvalues,
see Appendix C in \cite{Dolaptchiev2013} for the details.

\subsubsection{Homogenization}
\label{SEC_SMR}
The remaining step is the derivation of an effective equation for the
coarse variable \(\textbf{x}\) alone, using the homogenization
technique \cite{Majda2001,Pavliotis2008}, with terms taking SGS
effects into account.  The main assumption of the SMR is the presence
of distinct time scales in the considered variables. So far the model
is spatially separated into coarse and fine variables. This does not
necessarily imply a separation of the underlying time scales. However,
as it will be shown later, for the considered regime the separation in
space also induces a separation in time, with the resolved variable
\(\textbf{x}\) acting on a slower time scale than the SGS variable
\(\textbf{y}\). Hence a time-scale separation factor \(\epsilon \ll
1\) is introduced to characterize the different time scales associated
with the different terms on the right hand side of \eqref{OU-DNS_x}
and \eqref{OU-DNS_y}.  We replace \(b^x \rightarrow b^x/\epsilon\),
\(b^y \rightarrow b^y/\epsilon\), and \(\Lambda_{ij} y_j + \Sigma_i
\dot{W}_i \rightarrow \Lambda_{ij} y_j /\epsilon^2 + \Sigma_i
\dot{W}_i /\epsilon \), where then \(b^x\), \(b^y\), \(\Lambda_{ij}
y_j\), and \(\Sigma_i \dot{W}_i\) are all \(O(1)\), obtaining

\begin{align}
\dot{x}_i =& \varrho^x_i + a^x_i (\textbf{x}) + \frac{1}{\epsilon} b^{x}_i(\textbf{x},\textbf{y}) \;,
\\
\dot{y}_i =& \frac{1}{\epsilon} b^y_i(\textbf{x},\textbf{y}) + \frac{1}{\epsilon^2} \Lambda_{ij} y_j + \frac{1}{\epsilon}\Sigma_i \dot{W}_i  \;.
\label{Full eq. epsilon}
\end{align}

The above scaling implies that the bare truncation part \(\varrho^x + a^x(\textbf{x})\) acts on the slowest, the coupling terms \(b^x\) and \(b^y\) on a faster and the SGS self-interactions on the fastest time scale. In the following  the corresponding backward Fokker-Planck equation (FPE)
\begin{align}
\partial_t p =& L_3 p + \frac{1}{\epsilon} L_2 p + \frac{1}{\epsilon^2} L_1 p\, , \label{FPE full}
\end{align}
for the probability density function (PDF) \(p\) is considered, in the
limit of an infinite time scale separation \(\epsilon \rightarrow 0\).
The operators on the right hand side are defined as
\begin{align}
  L_3 =& - \left( \varrho^x_i + a^x_i (\textbf{x}) \right) \partial
  _{x_i}\, ,
  \\
  L_2 =& - b^{x}_i (\textbf{x},\textbf{y}) \partial_{x_i} - b^y_i
  (\textbf{x},\textbf{y}) \partial_{y_i} \, ,
  \\
  L_1 =& - \Lambda_{ij} y_j \partial_{y_i}
  -\frac{\Sigma_i^2}{2}\partial_{y_i}^2 \, .
\end{align}

Next, the PDF in   \eqref{FPE full} is expanded in terms of $\epsilon$: \(p = p^0 + \epsilon p^1 + \epsilon^2 p^2 + ...\), which leads to the following set of equations

\begin{align}
O(\epsilon ^{-2}): \quad 0 =& L_1 p^0 \;, \\
O(\epsilon ^{-1}): \quad 0 =& L_2 p^0 + L_1 p^1 \;, \\
O(\epsilon ^{0}): \quad p^0 =& L_3 p^0 + L_2  p^1 + L_1 p^2 \;.
\end{align}

The leading order \(O(\epsilon^{-2})\)-equation shows that \(p^0 \) is in the null space of \(L_1\) and therefore it does not depend on the fast variables: \(p^0 = p^0(\textbf{x})\). The \(O(\epsilon^{-1})\)-equation can be solved for  \(p^1\): $p^1 = - L_1^{-1}L_2p^0$ if the solvability condition
\begin{align}
 P L_2 p^0=0 \label{Solvability_cond}
\end{align}
is satisfied, where the projection operator $P$, projecting onto the
null space of $L_1$ is utilized. With this result the last
\(O(\epsilon^{0})\)-equation can be written as an effective FPE for
$p^0$ only

\begin{align}
\partial_{t} p^0 = L_3 p^0 - P L_2 L_1 ^{-1} L_2 p^0 \label{FPE reduced} \; .
\end{align}

This is the backward FPE of a low-resolution model, for the coarse
variables alone, which consists of the bare truncation part \(L_3
p^0\) and an SGS parameterization \(P L_2 L_1^{-1} L_2 p^0 \).
The null-space projection $P$ and inverse of the OU operator $L_1$ are
detailed in Appendix B. Using these, one finds that the stochastic
differential equation corresponding to the effective FPE \eqref{FPE
  reduced} can be written as
\begin{align}
dx_i = \left[ \varrho^x_i + a^x_i(\textbf{x}) + \beta_i(\textbf{x}) \right] dt + d\xi_{i}(\textbf{x}) \label{RSM} \;.
\end{align}
Here \(\beta_i\) represents the deterministic part and \(d\xi_i\) the
stochastic part of the SGS parameterization, containing both additive
and multiplicative noise terms. One finds that the deterministic
closure $\beta_i$ is
\begin{align}
  \beta_i =& \int_0^{\infty} d\tau \left\langle b^x_j(\textbf{x},\textbf{y}) \frac{\partial b^x_i (\textbf{x},\textbf{y}(\tau))}{\partial {x_j} } \right\rangle
      \nonumber\\ &
      + \langle\textbf{y}\textbf{y}^T\rangle^{-1}_{jm} \int_0^{\infty} d\tau\left\langle y_m b^y_j (\textbf{x},\textbf{y}) b^x_i (\textbf{x},\textbf{y}(\tau)) \right\rangle
  -\int_0^{\infty} d\tau \left\langle \frac{\partial b^y_j (\textbf{x},\textbf{y}) }{\partial y_j}b^x_i (\textbf{x},\textbf{y}(\tau))
 \right\rangle ,
\end{align}
where the expectations $\langle \cdot \rangle$ are taken over the OU statistics, and $\textbf{y}$ represents an OU trajectory with initial condition \( \textbf{y} = \textbf{y}(0)\). The stochastic closure $d\xi_i$ takes the form
\begin{align}
d \xi _i =
 & \sqrt{2} B_{ij} dW_j\, , \label{RSM-noise}
\end{align}
where the matrix elements $B_{ij}$ are obtained from the decomposition
\begin{align}
B_{ik} B_{jk} = \int_0 ^{\infty} d\tau \left\langle b^x_i(\textbf{x},\textbf{y}(0)) b^x_j(\textbf{x},\textbf{y}(\tau)) \right\rangle \;.
\label{eq_Bij}
\end{align}

With these results one can also show that back-transforming
\(b^x/\epsilon \rightarrow b^x\), \(b^y/\epsilon \rightarrow b^y\),
and \(\Lambda_{ij} y_j /\epsilon^2 + \Sigma_i \dot{W}_i /\epsilon
\rightarrow \Lambda_{ij} y_j + \Sigma_i \dot{W}_i\) leaves $\beta$ and
$d\xi$ unchanged so that (\ref{RSM}) is the desired low-resolution
model. 

Finally getting back to the specific case, (\ref{OU-DNS_x}) --
(\ref{eq_b_yi}) , Appendix C shows that the solvability condition
(\ref{Solvability_cond}) is satisfied. Moreover, inserting
(\ref{eq_b_xi}) and (\ref{eq_b_yi}) for $b^x$ and $b^y$ yields
\begin{align}
\beta _i = &
\left( L^{xy}_{mj} + B^{xxy}_{mlj} x_l \right)  B^{xxy} _{imk}  (C_S)_{jk}
\nonumber\\
&+
 \left( L_{mo} ^{yx} x_o + B_{mop} ^{yxx} x_o x_p \right)
 \left(  L_{ik} ^{xy} + B_{ijk} ^{xxy} x_j \right) \langle \textbf{y} \textbf{y}^T \rangle^{-1}_{mn}  (C_S)_{nk}
 \nonumber\\ &
+
B_{ijk} ^{xyy}  B_{mpo} ^{yxy} x_p  \langle \textbf{y} \textbf{y}^T \rangle^{-1}_{mn}
(C_T)_{onjk} \;,
\label{RSM-determiniistic}
\end{align}
where the tensors \(C_S\) and \(C_T\) are given by
 \begin{align}
 (C_S)_{jk}
& =
  \int_0^\infty d\tau \langle  y_j(0) y_k(\tau) \rangle \;,
  \\
   (C_T)_{onjk}
& =
 \int_0^\infty d\tau \left( \langle y_o(0) y_j(\tau)\rangle \langle y_k(\tau) y_n(0) \rangle + \langle y_o(0) y_k(\tau)\rangle \langle y_j(\tau) y_n(0) \rangle \right) \;.
 \end{align}
With this the decomposition (\ref{eq_Bij}) becomes
\begin{align}
B_{ik} B_{jk} =
B_{imn}^{xyy}(C_T)_{mnkl} B_{jkl}^{xyy} + \left(L^{xy}_{in} +  B^{xxy}_{ikn} x_k \right) (C_S)_{nm} \left(L^{xy}_{jm} +  B^{xxy}_{jlm} x_l \right) \;.
\end{align}

The prescription would be to perform this decomposition every time
step. However, this would be very expensive. Therefore we neglect
cross-correlations between the $d\xi_i$ in different cells and
approximate them by

\begin{align}
d\xi _i
\approx &  \sqrt{2 B_{ij} B_{ij}} dW_i \, ,\nonumber\\
=&
\sqrt{2 B_{ijn}^{xyy} (C_T)_{jnkl} B_{ikl}^{xyy} } dW^1_i
+
\sqrt{
2  \left(L^{xy}_{in} +  B^{xxy}_{ijn} x_j \right) (C_S)_{nk} \left(L^{xy}_{ik} +  B^{xxy}_{ilk} x_l \right) } dW^2_{i} ,
\label{RSM-effect-noise}
\end{align}
which we call effective stochastic forcing. In each coarse cell the
approximated stochastic term has thus the same variance as its exact
counterpart.  The stochastic term \eqref{RSM-effect-noise} consists of
an additive part, which acts on both variables,
and a multiplicative part, that acts only on \(HU\).  An important
feature of the SGS parameterization, with deterministic part
\eqref{RSM-determiniistic} and stochastic component
\eqref{RSM-effect-noise}, is that it couples a volume cell only to its
neighbors and next neighbors. This allows application of the approach
to large systems.

\section{Test case and model suite}
\label{SEC_Models}
For the validation of our approach we consider a stochastically forced
periodic shallow-water layer of horizontal extent \(L=10^4 \) km and
mean height \(\mathcal{H} = 10\) km. The diffusion constant is \(\nu =
10^5 \) km$^\text{2}$ day$^\text{-1}$ .  A large-scale stochastic forcing
\cite{chekhlov1995} is applied to the momentum equation
\begin{equation}
\pmb{\varrho}_I = \left( \begin{array}{c} 0 \\
\sum_{k=1} ^{3} \frac{ \mu \alpha_k }{\sqrt{k\Delta t}}
\text{cos} \left( 2\pi \left( \frac{k I n \Delta x}{L_x} + \psi_k \right) \right)\end{array} \right) \;.
\end{equation}
Normally distributed random numbers \(\alpha _k\) and \(\psi_k\) are
used, the amplitude parameter $\mu$ is $10^5$
{\(\text{km}^2 /
  \text{day}^{\frac{3}{2}}\)} and the forcing acts onto the leading
Fourier modes $1 \le k \le 3$. Various model set-ups have been chosen
as follows, a summary is given in Table \ref{t2}. In all
cases the integrations have been done over \(10^4 \text{days}\) with
\(10^3\) outputs per day.

\subsection{High-resolution simulations}

\subsubsection{DNS}
Reference is provided by direct numerical simulations, integrating
\eqref{DGL dicretised} with $N = 512$ volume cells. A 4th-order
Runge-Kutta-scheme is used, with a time step \(\Delta t = 10^{-4}\)
days.

\subsubsection{OU-DNS}

In two intermediate steps in the application of the SMR, first the
nonlinearities affected by SGS dynamics have been kept quadratic by
replacing $1/h \rightarrow 1/\mathcal{H}$, and then the SGS nonlinear
self-interactions have been replaced by an empirical OU process,
leading to the system \eqref{OU-DNS_x} and \eqref{OU-DNS_y}. Direct
integration of these equations, henceforth termed OU-DNS, thus appears
as a useful check of the validity of the SMR approach.  However,
directly using the OU parameters estimated from \eqref{OU-parameter_g}
and \eqref{OU-parameter_s} turned out not to be stable enough.
Therefore following \cite{Achatz1999} an additional
scale-selective damping has been supplemented to the OU drift in each
coarse cell. This has been done in the spectral representation of the
latter, see \eqref{DGL_fine_fourier}, by replacing

 \begin{align*}
 \Gamma \rightarrow \Gamma +
  \begin{pmatrix}
  \gamma & \ldots & 0 \\
   \vdots & \ddots & \vdots \\
   0 & \ldots & \gamma
   \end{pmatrix} \; ,
 && \text{with }
 \gamma = - \alpha
  \begin{pmatrix}
  1^2 & \ldots & 0 \\
  \vdots & \ddots &\vdots \\
  0 & \ldots & (n-1)^2
 \end{pmatrix}
 \in \mathbb{R}^{(n-1)\times(n-1)} \;.
 \end{align*}

 The diagonal matrix $\gamma$ represents damping of the Fourier modes
 inside each coarse cell with an amplitude proportional to the squared
 wave number, with here $\alpha = 90$ day$^{-1}$. As also in all
 other stochastic integrations outlined below, the time integration
 has been done by a split-step method with a 4th-order Runge-Kutta
 step for the deterministic part and an Euler-Mayurama step for the
 stochastic part. The time step is \(\Delta t = 10^{-4} \text{day} \),
 as in the deterministic DNS.

\subsection{Low-resolution simulations}

With the high-resolution simulations as reference we can validate the
SMR approach for providing an SGS parameterization for low-resolution
models that only use the coarse cells. We compare the performance of
this approach also to that of more simple purely empirical parameterizations.
Considered approaches are as follows where, unless otherwise stated,
the number of coarse cells employed was always $N_c=64$ with an averaging 
interval of $n=8$.

\subsubsection{Low-resolution simulations without SGS parameterization}

Two slightly different approaches have been chosen to obtain
low-resolution models.  The first is defined by the original
discretized equations \eqref{DGL dicretised}, but with a lower spatial
resolution of $N_c=N/n$. This is henceforth referred to as
\textit{low-resolution model (LRM)}. The second variant is the
\textit{bare-truncation model (BRT)} defined in
\eqref{Bare-Truncation}, with a resolution of $N_c$ as well. The
difference between BRT and LRM is the diffusion in the models. In the
BRT it is proportional to \(\nu/(n \Delta x^2)\), and in the LRM to
\(\nu/(n\Delta x)^2\), implying that the BRT has an effective
diffusion by a factor $n$ stronger, when compared to the LRM.

\subsubsection{Low-resolution simulations with SGS parameterization}

Three types of low-resolution simulations with 
stochastic SGS parameterization have been tested.

\paragraph{SMR parameterization.}

The low-resolution model \eqref{RSM} to be validated is the BRT
supplemented by the SMR SGS parameterization consisting of the
deterministic and stochastic components \eqref{RSM-determiniistic} and
\eqref{RSM-effect-noise}, it is referred to as BRT-SMR. For stability reasons
the BRT-SMR diffusivity had to be increased in
corresponding simulations to $\nu=2 \cdot 10^5$ km$^2$day$^{-1}$. The
time step employed was \(\Delta t = 2\cdot 10^{-5} \text{day}\).

\paragraph{Empirical OU parameterizations for BRT and LRM.}

As a quality measure for the SMR approach we also consider low-resolution simulations with an empirical OU SGS parameterization, denoted by BRT-OU or LRM-OU, depending on the low-resolution dynamical core used together with the empirical OU SGS parameterization. As in the SMR parameterization only coupling to neighbors and next neighbors is taken into account. The BRT-OU, for example, can be written as

\begin{align}
d{x}_i =& \left(\varrho^x_i + a^x_i (\textbf{x}) + \tilde{\Gamma}_{ij} \hat x^I_j \right )dt + \tilde{\sigma_i} dW_i\, .
\label{BRT-OU}
\end{align}

where the vector $\mathbf{\hat x}^I$ has 10 components and encompasses
the values of $H$ and $HU$ in the five coarse cell from $I-2$ up to
$I+2$, where $I$ is the cell index corresponding to the variable
$x_i$.  The OU parameters \(\tilde{\Gamma}\) and $\tilde \sigma$ have
been estimated using a standard maximum likelihood approach
\cite{Honerkamp1994}, yielding

\begin{align}
\label{tld_gamma}
\tilde{\Gamma}_{ij}
&=
\overline{b^x_i \hat x_k} \left(\overline{ \mathbf{\hat x} \mathbf{\hat x} ^T }^{-1}\right)_{kj}\, ,\\
\tilde{\sigma}_i^2 &= \Delta t \overline{ \left[ b^x_i - \tilde{\Gamma}_{ij} \hat x^I_j \right]^2 } \;,
\end{align}

with the superscript of $\mathbf{\hat x}^I$ suppressed in \eqref{tld_gamma}. For the LRM-OU the corresponding parameters have been determined in
the same manner. In contrast to the estimation of the OU processes in
the SMR by \eqref{OU-parameter_g} and \eqref{OU-parameter_s}, here the
integrated lagged covariance function could not be used, because the
SGS effects $b^x_i(\mathbf{x},\mathbf{y})$ as such do not satisfy a prognostic equation
dominated by an OU process. Replacing \eqref{OU-parameter_g} and
\eqref{OU-parameter_s} by maximum-likelihood estimates would have been
an option as well. Corresponding tests have shown a slightly
deteriorated performance, however.

\section{Results \label{SEC_Results}}

In the following we show step by step the essential results from our
various simulation experiments. Autocorrelations and spectra turned
out to be qualitatively similar for momentum and surface-height. We
therefore focus below on the latter. 

\subsection{DNS of the shallow-water layer\label{dns_results}
}
\begin{figure}[t!]
  \noindent\includegraphics[width=19pc, trim= 0.0cm 5.cm 0.1cm 5.4cm, clip ]{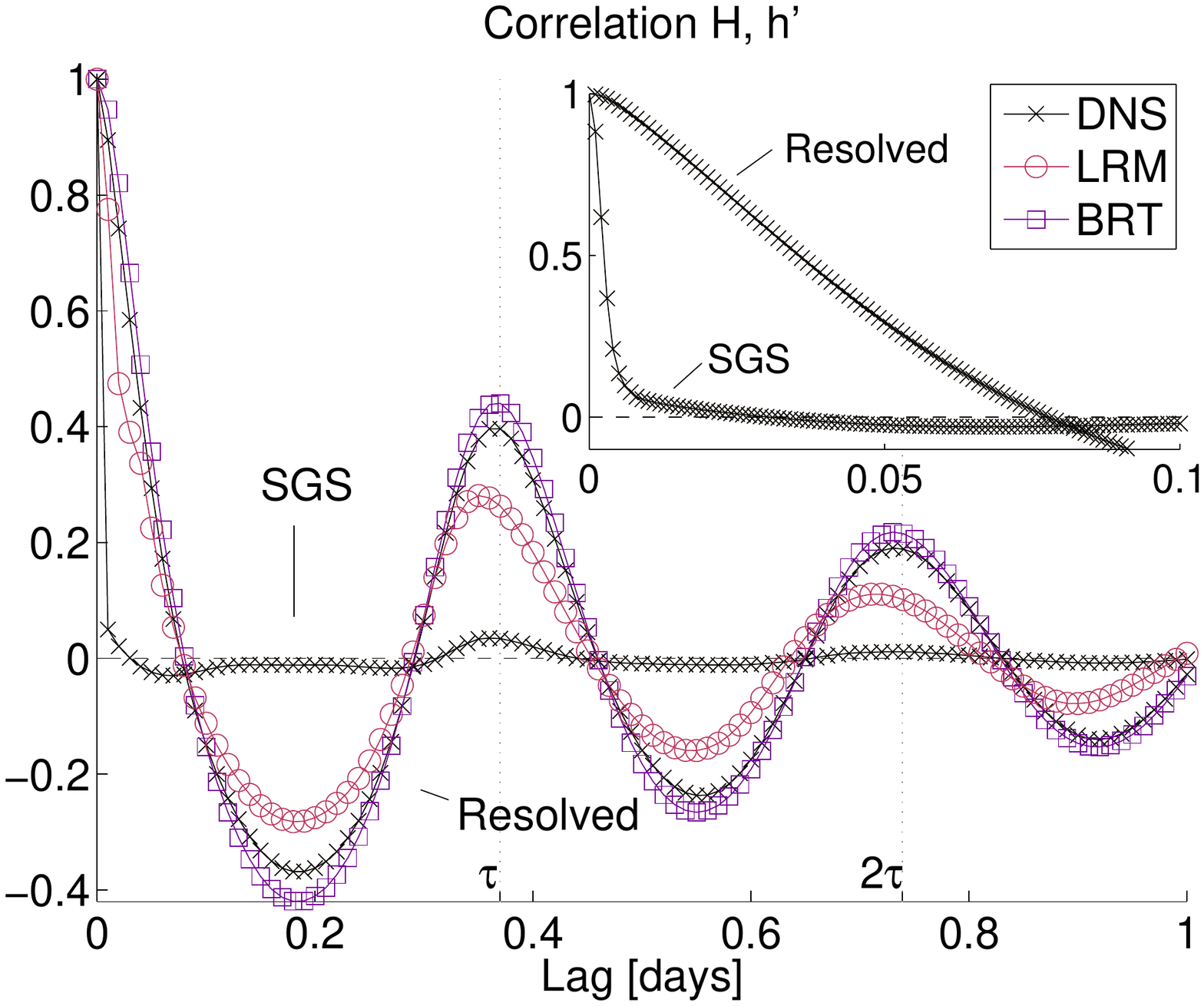}
  \noindent\includegraphics[width=19pc, trim= 0.0cm 5.cm 0.1cm 5.4cm, clip ]{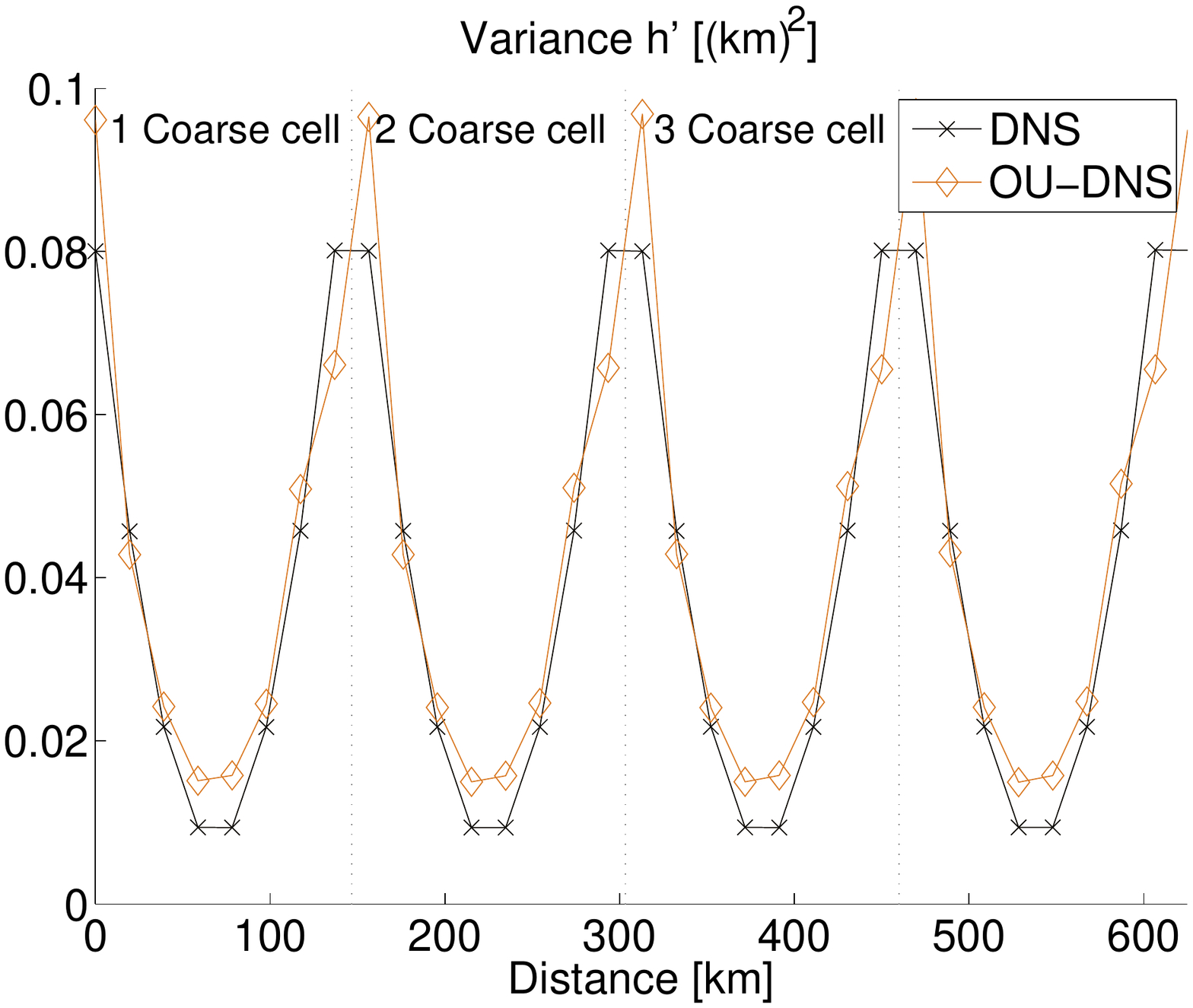}
  \caption{ Left: spatially averaged time autocorrelation of the
    resolved variable \(H\) and SGS variable \(h'\). Results from the
    DNS and two low-resolution simulations without SGS
    parameterizations (LRM and BRT). On the time axis the
    characteristic gravity wave time \(\tau\) is marked, see
    Sec.~\ref{SEC_Results}~\ref{dns_results}. In the upper right
    corner a shorter time interval is presented, which resolves the
    decay of $h'$. Right: the variance of the SGS variable \(h'\) for
    \(32\) fine grid points and an averaging interval \(n=8\). Results
    from the DNS and OU-DNS.}\label{fig-Corr-Var-DNS}
\end{figure}

Fig. \ref{fig-Corr-Var-DNS} (left) displays the time dependence of the
autocorrelation of the resolved variable \(H\) and of the SGS variable
\(h'\) in the DNS.  A slowly decaying oscillation is visible in the
autocorrelation of \(H\).  The period of this oscillation is nearly
equal to the time 
\( \tau = L/ \sqrt{g \mathcal{H}} \approx 0.37 \, \text{day} \), 
required for gravity waves to pass once
through the domain. This shows that the model has some intrinsic
dynamics and is not dominated by forcing and diffusion.
The autocorrelation of the SGS variable \(h'\) decays much faster to
zero than that of $H$. The large difference in the correlation time
between SGS and resolved variables indicates that the assumption of
time-scale separation between \(\textbf{x}\) and \(\textbf{y}\) is met
to a good agreement.

The spatial distribution of the variance of $h'$ is displayed in Fig.
\ref{fig-Corr-Var-DNS} (right). 
The variance is lowest in the middle of a coarse cell, and gradually
increases towards the cell boundaries. This spatial shape is explained
in Appendix E as being due to a spatially decreasing autocorrelation
of \(h\).

\begin{figure}[t]
  \noindent\includegraphics[width=19pc, trim= 0.0cm 5.cm 0.1cm 5.4cm, clip ]{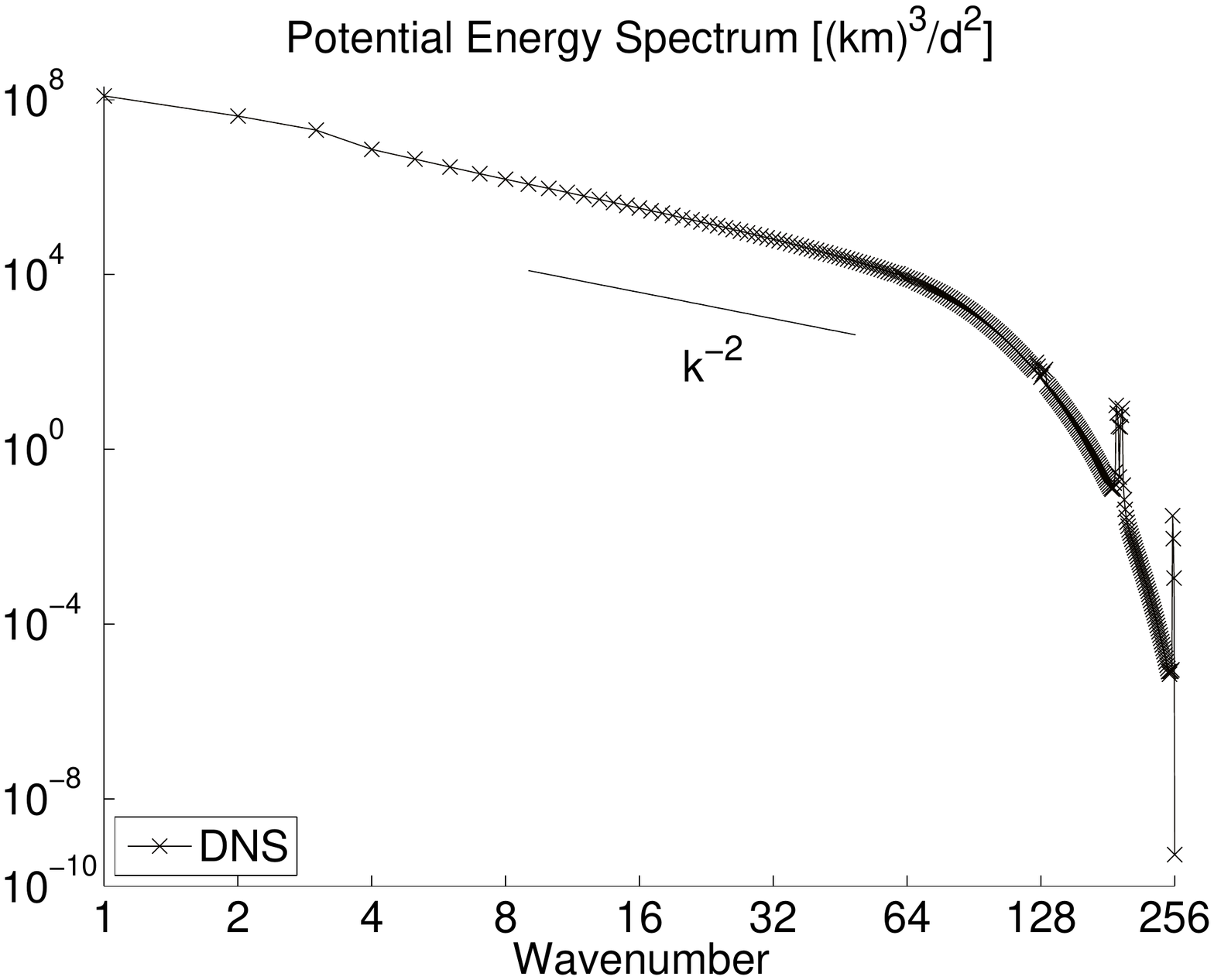}
  \noindent\includegraphics[width=19pc, trim= 0.0cm 5.cm 0.1cm 5.4cm, clip ]{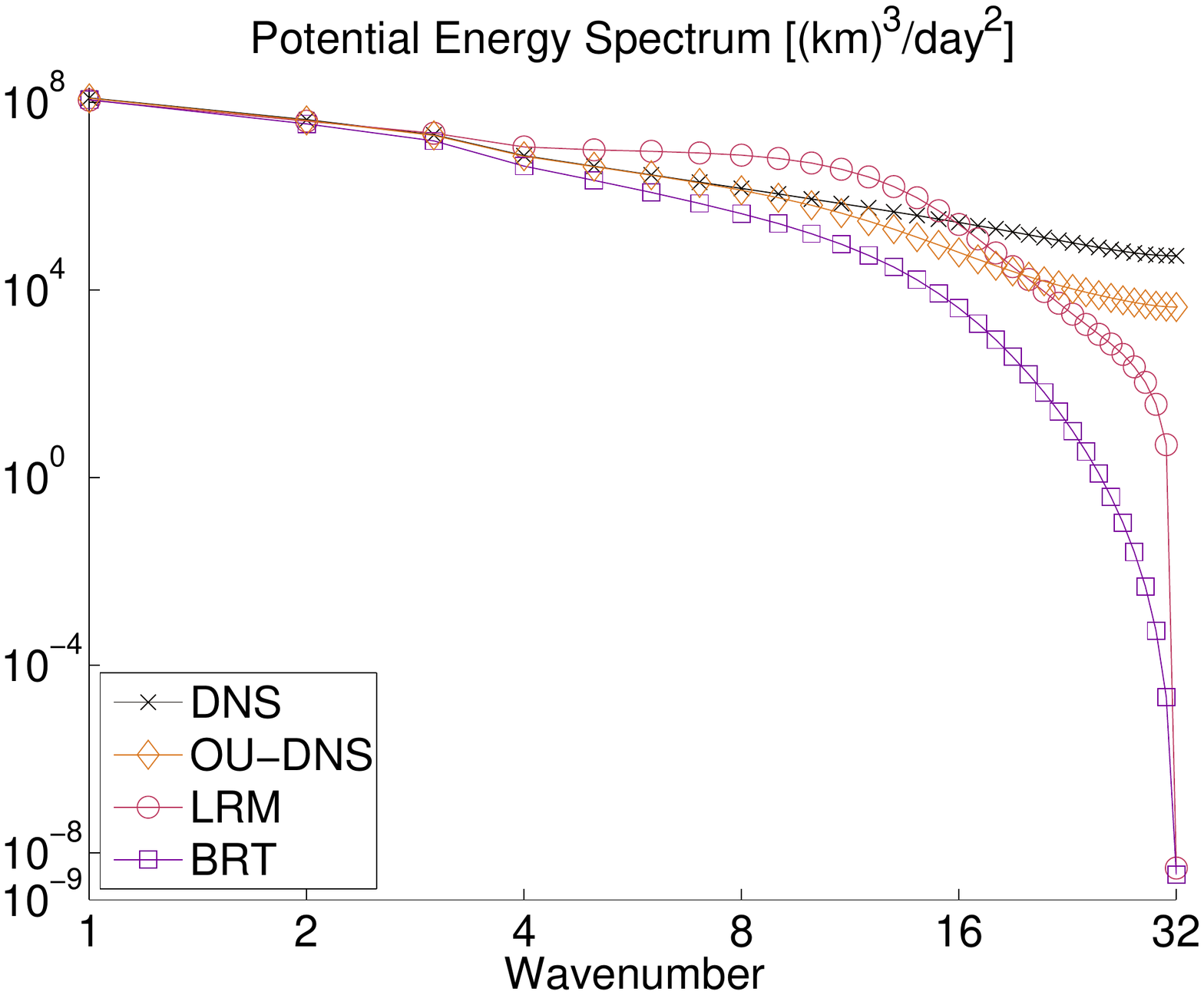}
  \caption{Left: the potential energy spectrum computed from $h$ in the DNS. Right: the potential energy spectrum computed from $H$ in the DNS, OU-DNS and two low-resolution simulations without SGS
    parameterizations (LRM and BRT).}\label{fig-Spec-DNS}
\end{figure}

The potential-energy spectrum from the DNS is displayed in the left panel of Fig. \ref{fig-Spec-DNS}. With
the considered forcing and diffusion parameters
one obtains an inertial range with spectral index 2
up to around wavenumber
\(kL/2\pi = 64\).
There is a small kink in the spectrum after wavenumber 3, due to the forcing acting only onto the first three modes.

The deviations from a Gaussian in the fourth order moments of $H$ and
$HU$ are less than $4\%$ and $2\%$, respectively. In addition we find
nearly vanishing odd moments (not shown). We conclude that the
statistics of the resolved variables are close to Gaussian.

\subsection{OU-DNS}
As described above, the replacement \(1/h \rightarrow 1/\mathcal{H}\)
in the SGS nonlinearities leads to the system \eqref{Full-h_x} and
\eqref{Full-h_z} with strictly quadratic nonlinearities, as required
for the application of the SMR method. Simulations with this model
reproduce the DNS data nearly perfectly (not shown).

Replacing the SGS self-interactions by an empirical OU process leads to the system \eqref{OU-DNS_x} and \eqref{OU-DNS_y}. The corresponding OU-DNS reproduces the correlations of the DNS with minor differences (not shown).
The energy spectrum from the OU-DNS, projected onto the coarse grid is
displayed in Fig.  \ref{fig-Spec-DNS} (right). It follows the DNS
spectrum for the first \(7\) wavenumbers and then drops below it. This
indicates a too strong damping at high wavenumbers, which seems to be
due to the introduction of the deterministic part in the OU-process.
The spatial variance of \(h'\) from the OU-DNS model is presented in Fig. \ref{fig-Corr-Var-DNS} (right). It follows with small deviations the structure from the DNS model.

\subsection{Low-resolution simulations without SGS parameterization}
Before considering results from low-resolution simulations with the
various SGS parameterizations, we first address low-resolution
simulations without any parameterizations, to provide a useful
reference. The time dependence of the autocorrelation function of $H$
from these simulations is shown in Fig. \ref{fig-Corr-Var-DNS} (left).
The amplitude of the oscillation of the auto-correlation from the LRM
simulations is significantly weaker than from the DNS and has a relative
error of \(6.3\%\), computed for time lags between $0$ and $1$ day.
The corresponding oscillation from the BRT simulation is slightly
stronger correlated with that from the DNS, with a relative error of
\(2.3\%\). The corresponding period
matches that from the DNS whereas that from the LRM simulation is
shorter.
Comparing the corresponding energy spectra in Fig. \ref{fig-Spec-DNS}
(right) one can see that the energy from the BRT simulation is overall
less than from the DNS, and that the spectrum is steeper. In contrast
to this, LRM simulations yield too much energy between wavenumbers 4
and 15, and too little at smaller scales. At all scales LRM
simulations yield more energy than the BRT simulations.

The relative errors in Table \ref{t1} show that the LRM simulation
significantly overestimates the all statistical moments, the fourth moment
in \(HU\) even by \(111.9\%\). The BRT simulation yields moments that
are too small, in the case of the fourth moment by \(29\%\).

We also verified numerically that in order to reproduce the spectra of
the first 32 wavenumbers in the DNS with $N=512$ grid points, it is
possible to perform low-resolution DNS with at least 256 spatial
points. However, further reducing the number of points in the
low-resolution DNS significantly corrupts the spectra of the first 32
wavenumbers. Thus, this demonstrates the need for SGS
parameterizations if one wants to reduce the spatial resolution beyond
$N=256$.




\subsection{Low-resolution simulations with SGS parameterization}

\begin{figure}[t!]
  \noindent\includegraphics[width=19pc, trim= 0.0cm 5.cm 0.1cm 5.4cm, clip ]{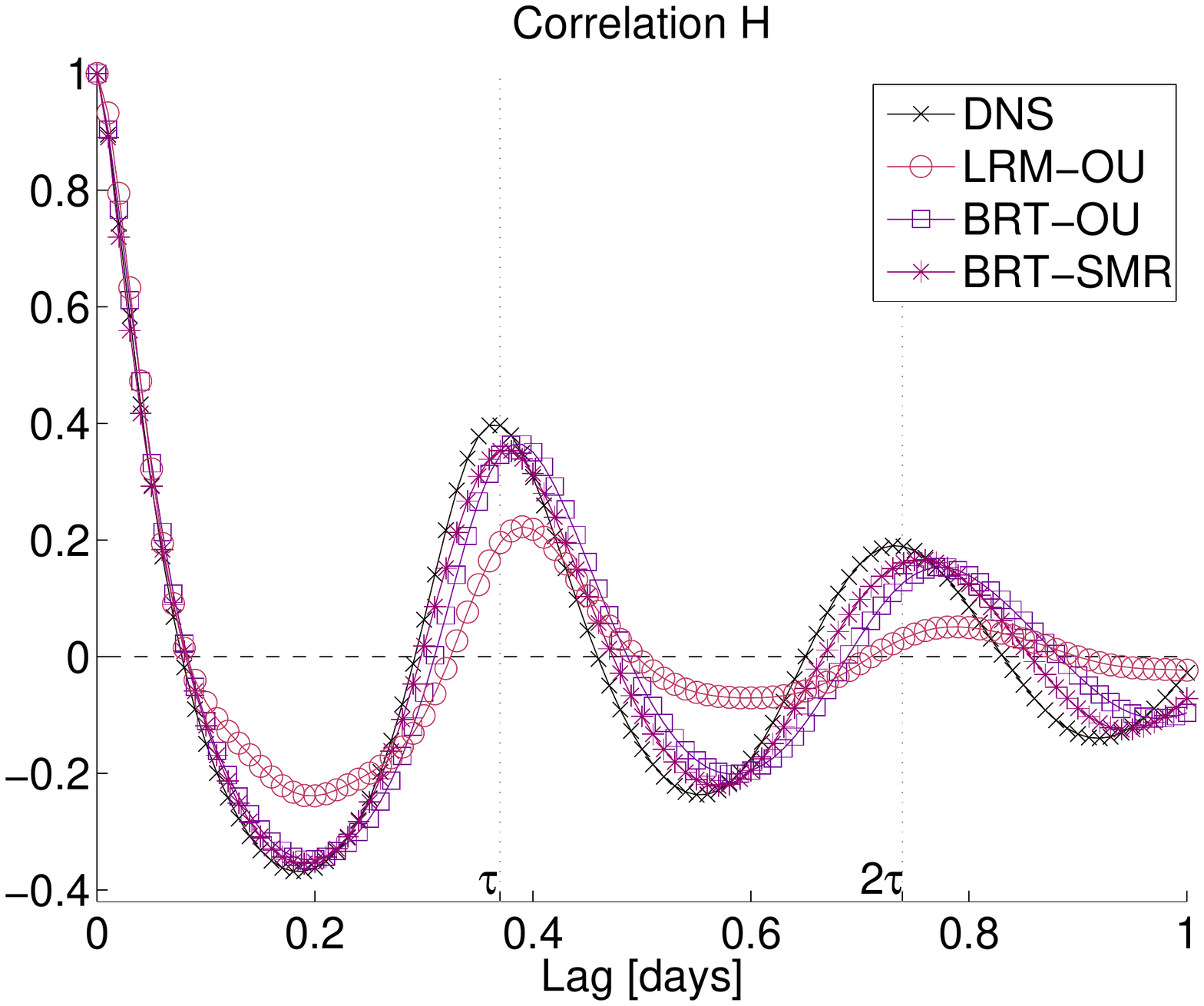}
  \noindent\includegraphics[width=19pc, trim= 0.0cm 5.cm 0.1cm 5.4cm, clip ]{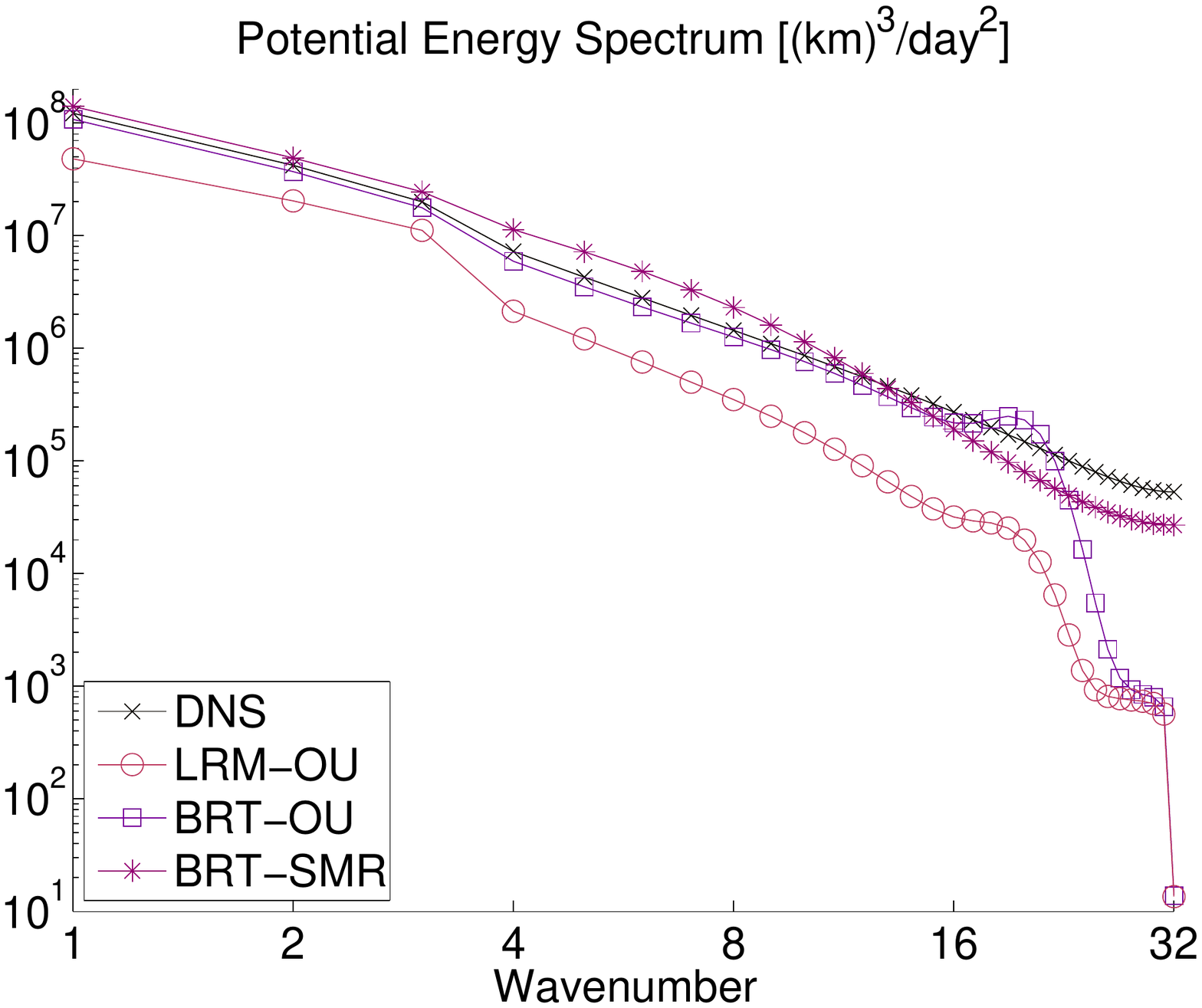}
  \caption{The spatially averaged time autocorrelation (left) and
    the potential energy spectrum (right) from DNS and three
    low-resolution simulations with SGS parameterizations  (LRM-OU,
    BRT-OU, BRT-SMR).} \label{fig-RSM}
\end{figure}

\textbf{Energy spectrum.}
The potential-energy spectra obtained from low-resolution simulations
with the different parameterizations are shown in Fig.  \ref{fig-RSM}
(right).  The overall qualitative behavior of the
shallow-water layer can be reproduced with the SMR parameterization
but there is too much energy in scales up to around wavenumber \(12\)
and too little energy in higher wavenumbers. On the other hand, 
the bare truncation model with empirical stochastic corrections (BRT-OU)
and the low-resolution model (LRM-OU) do not reproduce all the 
details of the spectra sufficiently accurately. In particular, 
the LRM-OU spectrum
contains significantly too little energy in all wave numbers. The
BRT-OU simulation can reproduce the true spectrum well in the first
\(15\) wavenumbers but fails completely at higher wavenumbers.

\textbf{Moments.}  The statistical moments from the various
simulations are summarized in Table \ref{t1}. In general all
  closures show high relative errors, the empirical OU
parameterizations underestimate and the SMR parameterization
overestimates the moments. Errors in the $HU$-moments are smallest for
the low-resolution simulation using the SMR parameterization. In the
$H$-moments BRT-SMR has lower errors than LRM-OU but larger than
BRT-OU. The BRT-SMR model overestimates the fluctuations in the
  coarse $H$-variable, which is consistent with the result for
  potential energy spectra discussed above. This implies that the SGS
  stochastic forcing representing the energy backscatter is too
  strong.

\textbf{Improvement of Energy Balance.}
To further improve the performance of our SGS parametrization using 
the stochastic mode reduction, we consider BRT-SMR model with reduced
SGS stochastic forcing. This is motivated by the fact that there are several assumptions in the SMR approach (e.g. time-scale separation, representing the
fast variables to the leading order by the OU process, 
polynomial form approximation of the interaction terms in the equation for momentum). Thus, we consider BRT-SMR models where 
the stochastic part in the SMR parameterization is
reduced by {40\%} or completely neglected. 
Stochastic terms in the SMR parametrization
represent the energy backscatter of small scales onto resolved large
scales. It has been recognized that proper modeling of this phenomena
is particularly important in the context of geophysical turbulence
(see e.g. \cite{Palmer2001, palmer2009, Berner2009}).  Therefore, we
study how well the SMR parametrization reproduces this process and
whether various approximations introduced in the context of applying
the SMR to the shallow water equation impose additional sensitivity of
the BRT-SMR model to the stochasticity of the closure.

\begin{figure}[t!]
  \noindent
\includegraphics[width=19pc, trim= 0.0cm 5.cm 0.1cm 5.4cm, clip ]{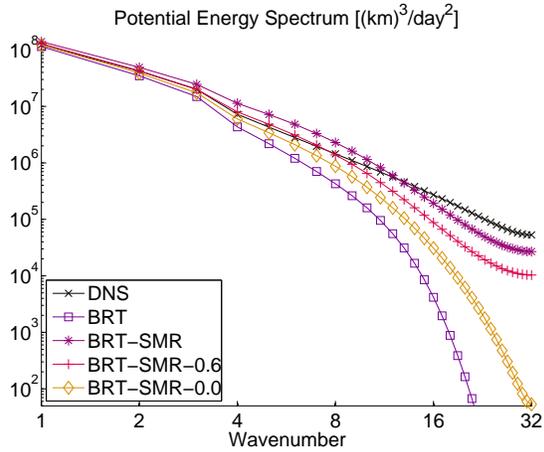}
\caption{The potential energy spectrum in DNS, BRT, BRT-SMR, BRT-SMR
  with a damped stochastic forcing \( d \xi \rightarrow 0.6 d \xi\)
  (BRT-SMR-0.6) and BRT-SMR with neglected stochastic forcing \(d\xi
  \rightarrow 0\) (BRT-SMR-0.0).}
\label{fig-RSM-stoch}
\end{figure}

The reduction of the stochastic part by \(40\%\) (defining BRT-SMR-0.6) significantly reduces the error in the variance of \(H\) to \(2.9\%\) and of \(HU\) to \(-5.2\%\), see Table \ref{t1}.
To avoid extensive tuning of the BRT-SMR model, we consider the uniform 
SGS noise reduction of both variables. Alternatively, SGS noises on 
$H$ and $HU$ can be reduced by a different percentage, thus further optimizing the
performance of the BRT-SMR model. 
With the choice of the 40\% reduction of SGS noise, the performance in
the energy spectrum can be improved for the first \(8\) wave numbers,
but for higher wave numbers the energy content drops, see Fig.
\ref{fig-RSM-stoch}. However, we wold like to emphasize that the
first 8 wavenumbers contain approximately \(97\%\) of the potential
energy.  From Fig.  \ref{fig-RSM-stoch} it is also visible that
already the deterministic part of the closure can significantly
improve the spectrum as compared to BRT.

\textbf{Correlation.}  The time autocorrelations from low-resolution
simulations (BRT or LRM) with either the empirical OU or
the SMR parameterization are depicted in Fig. \ref{fig-RSM} (left).
One can see that application of the SMR parameterizations leads
reproducing the autocorrelation from the DNS with small differences in
amplitude. The relative error of the correlation is \(3.4\%\).
Application of the OU parameterization in the LRM leads to simulations
with an oscillation in the autocorrelation that is too weak in
amplitude and exhibits a small phase shift, whereas use of the OU
approach in the BRT leads to simulations with an autocorrelation
similar to that obtained with the SMR parameterization.  The relative
error of the correlation is \(10.5\%\) for the LRM-OU simulation and
\(6.6\%\) for the BRT-OU simulation.  The SGS noise reduction in
  the BRT-SMR-0.6 model does not significantly affect the correlation
  function (not depicted for this model). The correlation function for
  the BRT-SMR-0.6 overlaps with the correlation function computed
  using the BRT-SMR. This can be intuitively understood since the
  correlation function in many stochastic models is determined
  primarily by the strength of the deterministic terms.

\subsection{Scale adaptivity}
The advantage of the SMR parameterization is that it can be adapted
easily to changes in the model setup, and in many situations it does
not have to be recalculated.  This has been investigated by
considering larger averaging intervals of \( \;n=16, \; 32\) ,
resulting in different spatial resolutions \(N_c = N/n = 16, \; 32\).
To adjust the SMR closure to
  the changed resolution, we use \eqref{RSM-determiniistic-scale} and
  \eqref{RSM-effect-noise-scale} from the Appendix D. Note that no
  re-determination of the model is necessary. In contrast to this, no
modification rule exists for the empirical closures in the BRT-OU and
the LRM-OU model. We keep those parameterizations unchanged for the
considered cases. Whereas the LRM-OU remains stable, the BRT-OU is
unstable in both cases.

\begin{figure}[t!]
  \noindent\includegraphics[width=19pc, trim= 0.1cm 5.cm 0.1cm 5.4cm, clip ]{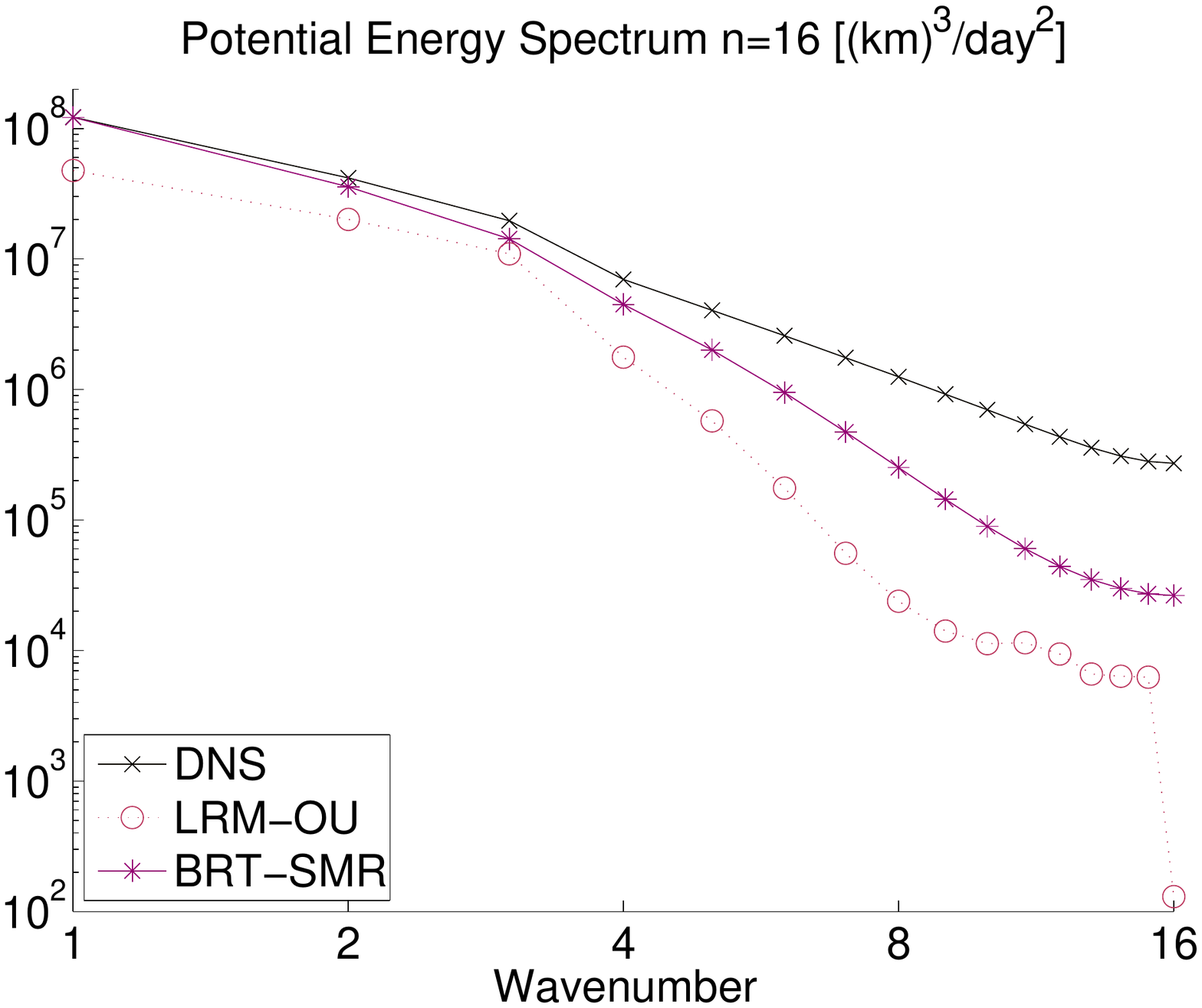}
  \noindent\includegraphics[width=19pc, trim= 0.1cm 5.cm 0.1cm 5.4cm, clip ]{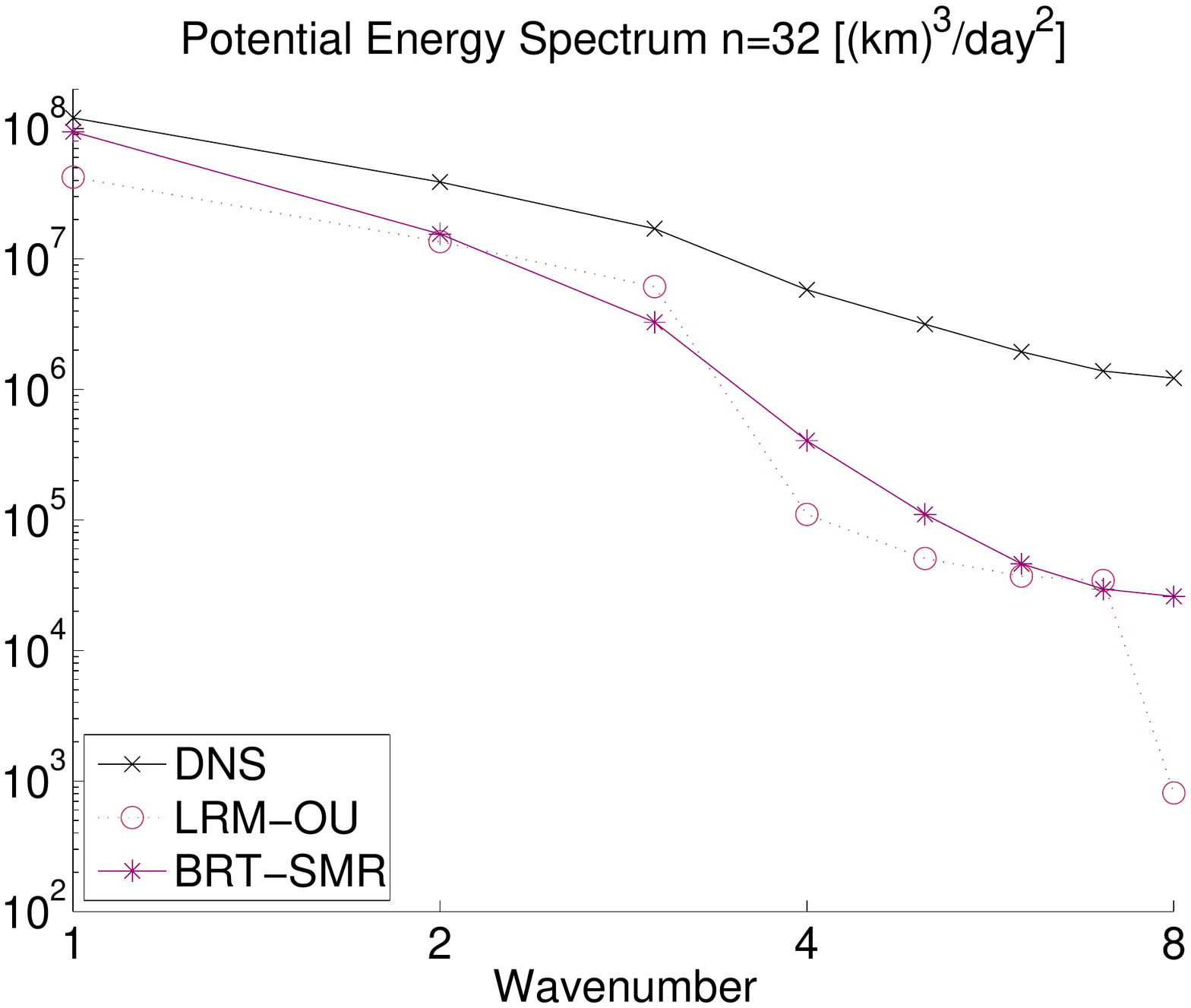}
  \caption{The Spectrum for an averaging interval of \(n=16\) (left)
    and $n=32$ (right) in DNS, LRM-OU and BRT-SMR. The simulations
    with LRM-OU and BRT-SMR have a resolution of $N_c=32$ (left) and
    $N_c=16$ (right).} \label{fig-AV}
\end{figure}

The potential energy spectra from integrations of the resulting stable
models are displayed in Fig. \ref{fig-AV}. For comparison the
corresponding DNS projection is shown as well. In both cases
integration of the low-resolution models with SGS parameterization
yield less energy in the resolved flow than the DNS. However,
application of the SMR SGS parameterization leads to better agreement
with the DNS, especially for $n=16$. Both low-resolution simulations
can capture the time correlation well (not shown).

\section{Conclusion \label{SEC_Conclusion}}

The applicability of subgrid-scale (SGS) parameterizations to a wide
range of parameters of a dynamical system such as the atmosphere, and
their ability to be easily used at different model
configurations, requires that they are based on
first principles as much as possible.  Stochastic mode reduction (SMR)
as suggested by \cite{Majda2001, Majda2002, Majda2003a}, i.e.
homogenization applied to a system with its nonlinear fast-variable
self-interactions replaced by an empirical Ornstein-Uhlenbeck (OU)
process, is a promising option in this direction. Geophysical
  applications of the SMR so far were performed always in spectral
  space \cite{Franzke2005,Franzke2006}. However, in many
  applications, such as ocean modeling or regional climate modeling,
  SGS parameterizations in physical space are required.  In order to
  construct such parameterization we use the local approach suggested
  by \cite{Dolaptchiev2013, Dolaptchiev2013a}, and tested within the
  framework of the Burgers equation.  A central
aspect of this approach is the discrimination, within a finite-volume
formulation of the high-resolution dynamics, between slowly varying
spatial averages, that are resolved explicitly, and more rapidly
varying deviations from those, that are to be parameterized.

As a next step towards the application of this technique to real
atmospheric flows, our work validates the applicability of the SMR in
the context of one-dimensional shallow-water (1DSW) flow. This
introduces two general features. (1) Gravity waves are included as
well as (2) high-order non-polynomial nonlinearities that generally
affect compressible flows. After the validation of the required
time-scale separation between local averages and small-scale flow, the
latter issue has been handled by replacing, in all dynamical terms
affecting or affected by the fast small-scale flow, the inverse of the
water-column height by the inverse of its global equilibrium value.
This limits the corresponding nonlinearities to quadratic. Further
replacing all small-scale self-interactions by an empirical
OU process yields a representation of the dynamics that allows model
simulations in rather good agreement with simulations of the
unmodified 1DSW equations. We could hence proceed and apply the
homogenization technique to obtain an explicit low-resolution model
for the local averages, with an SMR SGS parameterization of the
small-scale flow coupling only a small number of neighboring cells.

This model has been validated against data from high-resolution
simulations of 1DSW flow. It is shown that the SMR SGS
  parameterization improves the energy spectrum at the smaller
  resolved scales in comparison with both simulations without SGS
  parameterizations and simulations using an empirical OU SGS
  parameterization. In the error of some statistical moments
  no clear benefit of the SMR SGS parameterization is present.
  However, we demonstrated that the error can be considerably lowered
  by diminishing the stochasticity in the SMR closure. In particular,
  the variance error of the SMR SGS model can be reduced to \(2.9\%\)
  in \(H\). We also found that this comes along with less energy at
  high wavenumbers.  We conjecture that the performance of the
BRT-SMR model can be improved further by empirically adjusting the
coefficients of the SMR SGS parametrization.  Finally, we also show
that the closure can easily be adapted to changes in the model
parameters. This enables a scale awareness, which allows to utilize
the SMR SGS parameterization for different spatial resolutions and
leads to improvements compared to empirical SGS schemes.

In a related study within the framework of the Lorenz 96 model,
\cite{Vissio2017} have recently demonstrated parameter-awareness of a
parameterization derived using response theory \cite{wouters2012} by
changing the time-scale separation between resolved and unresolved
scales. How far this extends to our setting, where scale-adaptivity
is considered with regard to the number of resolved modes, remains to
be investigated. Moreover, as shown by \cite{wouters2016} for
comparatively simple models, the SGS scheme of \cite{wouters2012}
does outperform the SMR parameterization at smaller time lags, but on
longer time scales it converges to the SMR result in the limit of
infinite time scale separation. Still, it would be interesting to
extend the work of \cite{demaeyer2018} by comparing both approaches in
more complex applications. 

Motivated by the DIA closure of \cite{frederiksen_97,frederiksen_06} 
applied a stochastic modeling approach
  accounting for memory effects of the turbulent eddies and
  constructed SGS parameterization from a high-resolution
  simulation. The resulting SGS model is local in spectral space and
  includes linear eddy drain viscosity and stochastic backscatter
  viscosity. The same approach was successfully used by
  \cite{kitsios_12,kitsios_13} to construct scale-aware SGS parameterizations,
  which reproduce the spectra exactly. Interestingly, the SMR provides
  additional nonlinear deterministic correction terms and
  multiplicative noise terms. Such terms might become important in
  situations where effects due to topography, intermittency or
  large-scale flow are relevant. In addition, the scaling laws found
  by \cite{kitsios_12,kitsios_13} for the eddy viscosities suggest that similar
  scaling laws might be valid for the parameters of the OU process in
  Fourier space, used in the SMR approach. This might improve further
  the results on scale-adaptivity presented here.

Potentially an issue is that we had to increase diffusivity
in order to stabilize the low-resolution model with SMR SGS
parameterization. This is a well known issue with purely empirical SGS
parameterizations (e,g, \cite{Achatz1997}). In the present
semi-analytical approach it might be overcome by using the energy
conserving discretization of \cite{Fjordholm2011}. As pointed out by
\cite{Majda2009} there is a connection between energy conservation by
the discretized nonlinearities and the cubic damping term in the SMR
SGS parameterization. Indeed, in the studies of \cite{Franzke2005,
  Franzke2006, Dolaptchiev2013} the discrete treatment of the
nonlinear terms conserves energy and the resulting SMR SGS schemes are
stable.

Even from the present results, however, we conclude that it appears
worthwhile further moving towards the application of the SMR SGS
parameterizations to low-resolution simulations in general
compressible flows. Next step would be to
  increase the complexity by considering two-dimensional shallow-water
  flow and by including rotational effects. Such system contains
  dispersive inertial gravity waves as well as geostrophic balanced
  flow. One interesting question in this regard is if the effect of
  high-frequency, small-scale gravity waves on the large-scale gravity
  waves and geostrophic flow can be parameterized using the present
  local SMR approach.
%

%
\subsubsection*{Acknowledgments}
We want to thank the reviewers for their
    comments and suggestions which helped to improve the draft version
    of the manuscript. The code for the SGS parameterizations is
    available upon request. MZ, SD and IT thank the German Research
  Foundation (DFG) for partial support through grant DO 1819/1-1.  IT
  thanks for partial support by the grant ONR N00014-17-1-2845.  UA
  thanks DFG for partial support through grant AC 71/7-1.

\appendix

\section{Interaction coefficients}
To define the interaction coefficients in \eqref{Full-h_x}, \eqref{Full-h_z}, the following notation is used

\begin{align}
  \left(  
L_{lm}^{xz} z_m , B^{xxz}_{lmn}x_m z_n, B^{xzz}_{lmn}z_m z_n
  \right)  
&=
\begin{cases}
    \left(L^{Hz}, B^{Hxz}, B^{Hzz} \right) 
&\text{ if } x_l \text{ denotes } H_I\\
  \left(L^{HUz}, B^{HUxz}, B^{HUzz} \right) 
&\text{ if } x_l \text{ denotes } HU_I
\end{cases}\, ,\\
   \left(  
 L_{lm}^{zz} z_m , L_{lm}^{zx} x_m
   \right)  
&=\begin{cases}
    \left( L^{h'z}, L^{h'x} \right) 
&\text{ if } z_l \text{ denotes } h'_i\\
  \left( L^{hu'z}, L^{hu'x} \right) 
&\text{ if } z_l \text{ denotes } hu'_i
\end{cases}\, ,\\
   \left(  
B^{zxx}_{lmn}x_m x_n, B^{zxz}_{lmn}x_m z_n, B^{zzz}_{lmn}z_m z_n
   \right)  
&=\begin{cases}
    \left( B^{h'xx}, B^{h'xz}, B^{h'zz} \right) 
&\text{ if } z_l \text{ denotes } h'_i\\
  \left( B^{hu'xx}, B^{hu'xz}, B^{hu'zz} \right) 
&\text{ if } z_l \text{ denotes } hu'_i
\end{cases}\, ,
\end{align}

with \(I \in \{0,1,...,N_c-1\} \) and \(i \in \{0,1,...,N-1\}\). The linear interaction coefficients read

\begin{align}
\label{lhz}
L^{Hz} =&  -
\frac{1}{2n\Delta x }
\left( hu'_{n(I+1)} + hu'_{n(I+1)-1}  - hu'_{nI} - hu'_{nI-1} \right)
\\ &
 +
 \frac{\nu}{n\Delta x ^2} \left(  h'_{n(I+1)} - h'_{n(I+1)-1} - h'_{nI} + h'_{nI-1} \right)\nonumber \\
\label{lhuz}
 L^{HUz} =&
\frac{\nu}{n\Delta x ^2} \left(  hu'_{n(I+1)} - hu'_{n(I+1)-1} - hu'_{nI} + hu'_{nI-1} \right)\\
L^{h'x} =&
- \frac{1}{2\Delta x }
\left(HU_{I[i+1]} - HU_{I[i-1]} \right) +  \frac{\nu}{\Delta x ^2} \left(  H_{I[i-1]} - 2 H_{I[i]} + H_{I[i+1]} \right) \\
&+ \frac{1}{2n\Delta x }
\left(HU_{I[i]+1} - HU_{I[i]-1} \right) -
 \frac{\nu}{n\Delta x ^2} \left(  H_{I[i]+1} - 2 H_{I[i]} + H_{I[i]-1} \right)
 \\
L^{hu'x} =&
 \frac{\nu}{\Delta x ^2} \left(  HU_{I[i-1]} - 2 HU_{I[i]} + HU_{I[i+1]} \right)  \\
& - \frac{\nu}{n\Delta x ^2} \left(  HU_{I[i]+1} - 2 HU_{I[i]} + HU_{I[i]-1} \right)\\
\label{lhpz}
L^{h'z} =& -
\frac{1}{2\Delta x }
\left(hu'_{i+1} - hu'_{i-1} \right) + \frac{\nu}{\Delta x ^2} \left( h'_{i-1} - 2 h'_{i} + h'_{i+1}  \right) - L^{Hz}
\\
\label{lhupz}
L^{hu'z} =&
  \frac{\nu}{\Delta x ^2} \left( hu'_{i-1} - 2 hu'_{i} + hu'_{i+1}  \right) - L^{HUz}\, ,
\end{align}

where in \eqref{lhpz}, \eqref{lhupz} the terms $L^{Hz}$ and $L^{HUz}$ are given in \eqref{lhz}, \eqref{lhuz}, but with the index $I$ replaced by $I[i]$. The nonlinear interaction coefficients  read

\begin{align}
B^{Hxz} &= B^{Hzz} = B^{h'xx} = B^{h'xz} = B^{h'zz} = 0\\
\label{bhuxz}
B^{HUxz} &=
-\frac{1}{n\Delta x}
\Bigg[
\frac{1}{\mathcal{H}}
\Big(
HU_{I+1} hu'_{n(I+1)} + HU_{I}hu'_{n(I+1)-1} \\
&- HU_{I}hu'_{nI} - HU_{I-1}hu'_{nI-1}
\Big)
 \nonumber\\ &
+ \frac{g}{2} \left( H_{I+1}h'_{n(I+1)} + H_{I}h'_{n(I+1)-1} - H_{I}h'_{nI} - H_{I-1}h'_{nI-1} \right)
\Bigg]\nonumber \\
\label{bhuzz}
B^{HUzz}
=& -\frac{1}{2n\Delta x}
\Bigg[
\frac{1}{\mathcal{H}} \left( (hu'_{n(I+1)})^2 + (hu'_{n(I+1)-1})^2 - (hu'_{nI})^2 - (hu'_{nI-1})^2\right)
\\ & +
 \frac{g}{2} \left( (h'_{n(I+1)})^2 + (h'_{n(I+1)-1})^2 - (h'_{nI})^2 - (h'_{nI-1})^2 \right)
\Bigg]
\nonumber \\
B^{hu'xx}
=& -\frac{1}{2\Delta x} \left( \frac{1}{\mathcal{H}} \left( HU_{I[i+1]}^2 - HU_{I[i-1]}^2 \right) + \frac{g}{2} \left( H_{I[i+1]}^2 - H_{I[i-1]}^2 \right)\right) \\
&+\frac{1}{2n\Delta x}
\left[ \frac{1}{\mathcal{H}} \left( HU_{I[i]+1}^2 - HU_{I[i]-1}^2\right)+
 \frac{g}{2} \left( H_{I[i]+1}^2 - H_{I[i]-1}^2 \right)
\right]
\\
B^{hu'xz}
=&
-\frac{1}{\Delta x}
\left[ \frac{1}{\mathcal{H}} \left( HU_{I[i+1]}hu'_{i+1} - HU_{I[i-1]}hu'_{i-1}  \right) + \frac{g}{2} \left( H_{I[i+1]}h'_{i+1} - H_{I[i-1]}h'_{i-1}  \right)
\right]
\nonumber\\
\label{bhupxz}
&- B^{HUxz}
\\
\label{bhupzz}
B^{hu'zz} =& -\frac{1}{2\Delta x} \left( \frac{1}{\mathcal{H}} \left( (hu'_{i+1})^2 - (hu'_{i-1})^2\right) + \frac{g}{2} \left( (h'_{i+1})^2 - (hu'_{i-1})^2 \right)\right) - B^{HUzz}\, ,
\end{align}

where in \eqref{bhupxz}, \eqref{bhupzz} the terms $B^{HUxz}$ and $B^{HUzz}$ are given in \eqref{bhuxz}, \eqref{bhuzz}, but with the index $I$ replaced by $I[i]$.

\section{Null-space projection and inverse of the OU backward Fokker-Planck operator}
To determine the projection operator \(P\) and the inverse operator \(L_1^{-1}\)
one can consider an auxiliary process described by the backward FPE \(\partial_t \chi = L_1 \chi \) with the conditional PDF \(\chi(\tilde{\textbf{y}}, 0 \mid \textbf{y}, \tau)\).  The invariant measure of this process \(p_s(\tilde{\textbf{y}}) = \underset{\tau \rightarrow -\infty}{\text{lim}} \chi(\tilde{\textbf{y}}, 0 \mid \textbf{y},\tau)\) defines the projection operator

\begin{align}
  \left( P \thinspace g\right)(\textbf{x}) = \int d\textbf{y} \;
  g(\textbf{x},\textbf{y}) p_s (\textbf{y}) = \langle
  g(\textbf{x},\textbf{y}) \rangle\;,
\end{align}

with the expectation \(\langle \cdot \rangle\) of \(g(\textbf{x},\textbf{y})\) with respect to the invariant measure \(p_s\). The inverse operator \(L_1^{-1}\) applied onto a function \(f(\textbf{x},\textbf{y})\) is found to be
given by

\begin{align}
  \left( L_1 ^{-1} \thinspace f \right)(\textbf{x},\textbf{y})
  = \int_0 ^{\infty} d\tau \int d\textbf{y} \;
  f(\textbf{x},\tilde{\textbf{y}})\chi(\tilde{\textbf{y}},\tau \mid \textbf{y},0) \;.
\label{p_opr}
\end{align}

Both operators applied consecutively yield

\begin{align}
  \left( P g L_1 ^{-1} \thinspace f \right)
  (\textbf{x})
  =& \int d\textbf{y} \; g(\textbf{x},\textbf{y}) p_s (\textbf{y})
  \int _0 ^{\infty} d\tau \int d\tilde{\textbf{y}} \;
  f(\textbf{x},\tilde{\textbf{y}})\chi(\tilde{\textbf{y}},
  \tau \mid \textbf{y}, 0)
  \nonumber\\
  =& \int _0 ^{\infty} d\tau \int d\textbf{y} \;
  g(\textbf{x},\textbf{y}) p_s (\textbf{y})\int d\tilde{\textbf{y}} \;
  f(\textbf{x},\tilde{\textbf{y}})\chi(\tilde{\textbf{y}},
  \tau \mid \textbf{y}, 0)
  \nonumber\\
  =& \int _0 ^{\infty} d\tau \; \left\langle
    g(\textbf{x},\textbf{y})
    f(\textbf{x},\tilde{\textbf{y}}(\tau)) \right\rangle\, ,
\end{align}

where in the lagged covariance in the last line $\tilde{\textbf{y}}(\tau)$ is understood to be an OU trajectory with initial condition $\tilde{\textbf{y}}(0) = \textbf{y}$.
%

\section{Solvability condition
} The solvability condition \eqref{Solvability_cond} can be rewritten
to \(P b^x_i \partial_{x_i} p^0=0\) since \(p^0=p^0(\textbf{x})\).
It is fulfilled if the even stronger condition \(P b^x_i = 0\)
holds. This is the case here.  Using \eqref{eq_b_xi}, defining

\begin{equation}
  \kappa_{jk} = R_{jl} R_{km} \langle y_l y_m \rangle
\end{equation}

and inserting the transformation \(\textbf{z} = R \textbf{y}\),
one obtains with $\langle y_i \rangle = 0$
\begin{align}
P b^x_i
= B^{xyy}_{ijk} \langle y_j y_k \rangle
= B^{xzz}_{ijk} \kappa_{jk}
= B^{xzz}_{ijj} \kappa_{jj} \;.
\end{align}

We have used in the last step that \(B^{xzz}_{ijk} = B^{xzz}_{ijk}
\delta_{jk}\), as can be seen from Appendix A.  Since \(B^{xzz}_{ijj}
\kappa_{jj}\) is a difference between fluxes at the right and at the
left of a cell, the total sum over \(i\) vanishes
\begin{align}
\sum _{i} B^{xzz}_{ijj} \kappa_{jj}  =& 0 \;.
\end{align}
 The homogeneity of $\langle y_l y_m \rangle$ implies that each element in the sum is identical and thus must vanish. This proofs the solvability condition.

\section{The SMR SGS parameterization for changed resolution}
The SMR SGS parameterization can be adapted to different coarse grid
resolutions without any recalculation. This can be achieved by
collecting interaction coefficients proportional to different powers
of $n$. For example the linear coefficients \(L^{yx}\) can be written
as \(L^{yx} = \tilde{L}^{yx} + \frac{1}{n} \hat{L}^{yx}\), where
$\hat{L}^{yx}$ results from all terms in $L^{yx}$ multiplied by \(
\frac{1}{n}\) and $\tilde{L}^{yx}$ from terms independent of $n$.
Similarly all other coefficients can be split in this way, in
particular we have \(L^{xy} = \frac{1}{n} \hat{L}^{xy}\), \(B^{xxy} =
\frac{1}{n}\hat{B}^{xxy}\), etc..  This separation leads to the
deterministic closure
\begin{align}
\beta _i = &
\left( \frac{1}{n}\hat{L}^{xy}_{mj} + \frac{1}{n}\hat{B}^{xxy}_{mlj} x_l \right)  \frac{1}{n}\hat{B}^{xxy} _{imk}  (C_S)_{jk}
\nonumber\\
&+
 \left( \left[\tilde{L}_{mo} ^{yx} + \frac{1}{n}\hat{L}_{mo} ^{yx})\right] x_o + \left[\tilde{B}_{mop} ^{yxx} + \frac{1}{n}\hat{B}_{mop} ^{yxx}\right] x_o x_p \right)
 \frac{1}{n}\left(  \hat{L}_{ik} ^{xy} + \hat{B}_{ijk} ^{xxy} x_j \right) \langle \textbf{y} \textbf{y}^T \rangle^{-1}_{mn}  (C_S)_{nk}
 \nonumber\\ &
+
\frac{1}{n} \hat{B}_{ijk} ^{xyy}  \left[\tilde{B}_{mpo} ^{yxy} +  \frac{1}{n} \hat{B}_{mpo} ^{yxy}\right] x_p  \langle \textbf{y} \textbf{y}^T \rangle^{-1}_{mn}
(C_T)_{onjk} 
\\
=&
\frac{1}{n} \tilde{\beta}_i + \frac{1}{n^2} \hat{\beta}_i \label{RSM-determiniistic-scale}
\end{align}
where in the last steps the results are summarized with respect to the power of \(n\). The effective stochastic closure analogously yields
\begin{align}
d\xi _i
\approx &  
\frac{1}{n} \sqrt{2 \hat{B}_{ijn}^{xyy} (C_T)_{jnkl} \hat{B}_{ikl}^{xyy} } dW^1_i
+
\frac{1}{n} \sqrt{
2  \left(\hat{L}^{xy}_{in} +  \hat{B}^{xxy}_{ijn} x_j \right) (C_S)_{nk} \left(\hat{L}^{xy}_{ik} +  \hat{B}^{xxy}_{ilk} x_l \right) } dW^2_{i}
\\
=& \frac{1}{n} d\hat{\xi} _i \label{RSM-effect-noise-scale}
\, .
\end{align}

\section{Spatial shape of the SGS variance}
In continuous space the surface-height mean in the first coarse cell, with length $L_c$, and the corresponding SGS deviations can be written as

\begin{align}
H
= \frac{1}{L_c} \int_0 ^{L_c}
h(x)
\; d
x
\;,
&&
h'(x) = h(x) - H
\;.
\end{align}

This leads to spatial dependence in the variance of $h'$

\begin{align}
\overline{ h'(x)^2 } - \overline{ h' }^2
=
\overline{ h(x)^2} - 2 \overline{h(x)  H } + \overline{  H^2} - \overline{ h' }^2
\;, \label{analyt-varaince-z}
\end{align}

Due to spatial homogeneity of $h(x)$ and $\overline{h'}=0$, only the
second term can be assumed to be spatially dependent. Now by assuming
an exponentially decaying spatial correlation \(\overline{ h(x) h(x')
} \sim \text{exp} \left(-\alpha |x-x'| \right)\), with a decay rate
$\alpha > 0$, the middle term becomes

\begin{align}
\overline{ h(x)
H
}
= &
\frac{1}{L_c} \int_0 ^{L_c} \overline{ 
h(x)   h(x')
}  \; dx'
\nonumber\\
\sim &
\frac{1}{L_c} \left\lbrace \int_0 ^{x}  \text{exp} \left[\alpha(x'-x) \right]\; dx' + \int_x ^{L_c}  \text{exp} \left[\alpha(x-x') \right]  \; dx' \right\rbrace
\nonumber\\
=&
\frac{1}{\alpha L_c} \left\lbrace 2 - \text{e}^{-\alpha x} - \text{e}^{\alpha(x-L_c)}   \right\rbrace
\nonumber\\
=&
\frac{2}{\alpha L_c} \left\lbrace 1 - \text{e}^{-\frac{\alpha L_c}{2}} \text{cosh}\left[ \alpha\left( x-\frac{L_c}{2}\right)\right] \right\rbrace \;,
\end{align}

describing the characteristic U-shape of the fine variable variance in
the first coarse cell \(x \in [0,L_c]\).  The same considerations hold
for all other coarse cells.



\begin{thebibliography}{10}

\bibitem{Achatz1997}
U~Achatz and G~Schmitz.
\newblock {On the Closure Problem in the Reduction of Complex Atmospheric
  Models by PIPs and EOFs: A Comparison for the Case of a Two-Layer Model with
  Zonally Symmetric Forcing}.
\newblock {\em Journal of the Atmospheric Sciences}, 54(20):2452--2474, 1997.

\bibitem{Achatz1999}
Ulrich Achatz and Grant Branstator.
\newblock {A Two-Layer Model with Empirical Linear Corrections and Reduced
  Order for Studies of Internal Climate Variability}.
\newblock {\em Journal of the Atmospheric Sciences}, 56(17):3140--3160, 1999.

\bibitem{Achatz2013}
Ulrich Achatz, Ulrike L{\"{o}}bl, Stamen~I. Dolaptchiev, and Andrey Gritsun.
\newblock {Fluctuation–Dissipation Supplemented by Nonlinearity: A
  Climate-Dependent Subgrid-Scale Parameterization in Low-Order Climate
  Models}.
\newblock {\em Journal of the Atmospheric Sciences}, 70(6):1833--1846, jun
  2013.

\bibitem{Arnold2003}
Ludwig Arnold, Peter Imkeller, and Yonghui Wu.
\newblock {\em {Reduction of deterministic coupled atmosphere-ocean models to
  stochastic ocean models: A numerical case study of the Lorenz-Maas system}},
  volume~18.
\newblock 2003.

\bibitem{Berner2009}
J.~Berner, G.~J. Shutts, M.~Leutbecher, and T.~N. Palmer.
\newblock {A Spectral Stochastic Kinetic Energy Backscatter Scheme and Its
  Impact on Flow-Dependent Predictability in the ECMWF Ensemble Prediction
  System}.
\newblock {\em Journal of the Atmospheric Sciences}, 66(3):603--626, 2009.

\bibitem{Berner2017}
Judith Berner, Ulrich Achatz, Lauriane Batt{\'{e}}, Lisa Bengtsson, Alvaro {De
  La C{\'{a}}mara}, Hannah~M. Christensen, Matteo Colangeli, Danielle~R.B.
  Coleman, Daaaan Crommelin, Stamen~I. Dolaptchiev, Christian~L.E. Franzke,
  Petra Friederichs, Peter Imkeller, Heikki J{\"{a}}rvinen, Stephan Juricke,
  Vassili Kitsios, Fran{\c{c}}ois Lott, Valerio Lucarini, Salil Mahajaajaajan,
  Timothy~N. Palmer, C{\'{e}}cile Penland, Mirjajana Sakradzijaja, Jin~Song
  {Von Storch}, Antje Weisheimer, Michael Weniger, Paul~D. Williams, and
  Jun~Ichi Yano.
\newblock {Stochastic parameterization toward a new view of weather and climate
  models}.
\newblock {\em Bulletin of the American Meteorological Society},
  98(3):565--587, 2017.

\bibitem{Buizza1999}
R~Buizza, M~Miller, and TN~Palmer.
\newblock {Stochastic representation of model uncertainties in the ECMWF
  Ensemble Prediction System}.
\newblock {\em Quarterly Journal of the Royal Meteorological Society},
  125(560):2887--2908, oct 1999.

\bibitem{chekhlov1995}
Alexei Chekhlov and Victor Yakhot.
\newblock {Kolmogorov turbulence in a random-force-driven Burgers equation:
  Anomalous scaling and probability density functions}.
\newblock {\em Physical Review E}, 52(5):5681--5684, 1995.

\bibitem{delsole2004}
Timothy Delsole.
\newblock {Stochastic models of quasigeostrophic turbulence}.
\newblock {\em Surveys in Geophysics}, 25(2):107--149, 2004.

\bibitem{demaeyer2018}
Jonathan Demaeyer and S~Vannitsem.
\newblock Comparison of stochastic parameterizations in the framework of a
  coupled ocean-atmosphere model.
\newblock 01 2018.

\bibitem{Demaeyer2017b}
Jonathan Demaeyer and St{\'{e}}phane Vannitsem.
\newblock {Stochastic parametrization of subgrid-scale processes in coupled
  ocean-atmosphere systems: benefits and limitations of response theory}.
\newblock {\em Quarterly Journal of the Royal Meteorological Society},
  143(703):881--896, jan 2017.

\bibitem{Dolaptchiev2013}
S.~I. Dolaptchiev, U.~Achatz, and I.~Timofeyev.
\newblock {Stochastic closure for local averages in the finite-difference
  discretization of the forced Burgers equation}.
\newblock {\em Theoretical and Computational Fluid Dynamics}, 27(3-4):297--317,
  2013.

\bibitem{Dolaptchiev2013a}
Stamen~I. Dolaptchiev, Ilya Timofeyev, and Ulrich Achatz.
\newblock {Subgrid-scale closure for the inviscid Burgers-Hopf equation}.
\newblock {\em Communications in Mathematical Sciences}, 11(3):757--777, 2013.

\bibitem{Fjordholm2011}
Ulrik~S. Fjordholm, Siddhartha Mishra, and Eitan Tadmor.
\newblock {Well-balanced and energy stable schemes for the shallow water
  equations with discontinuous topography}.
\newblock {\em Journal of Computational Physics}, 230(14):5587--5609, 2011.

\bibitem{Franzke2006}
Christian Franzke and Andrew~J Majda.
\newblock {Low-Order Stochastic Mode Reduction for a Prototype Atmospheric
  GCM}.
\newblock {\em Journal of the Atmospheric Sciences}, 63(2):457--479, 2006.

\bibitem{Franzke2005}
Christian Franzke, Andrew~J Majda, and Eric Vanden-Eijnden.
\newblock {Low-Order Stochastic Mode Reduction for a Realistic Barotropic Model
  Climate}.
\newblock {\em Journal of the Atmospheric Sciences}, 62(6):1722--1745, jun
  2005.

\bibitem{frederiksen_97}
J.~S. {Frederiksen} and A.~G. {Davies}.
\newblock {Eddy Viscosity and Stochastic Backscatter Parameterizations on the
  Sphere for Atmospheric Circulation Models.}
\newblock {\em Journal of Atmospheric Sciences}, 54:2475--2492, October 1997.

\bibitem{frederiksen_03}
J.~S. Frederiksen, M.~R. Dix, and A.~G. Davies.
\newblock The effects of closure-based eddy diffusion on the climate and
  spectra of a gcm.
\newblock {\em Tellus A}, 55(1):31--44, 2003.

\bibitem{frederiksen_06}
J.~S. {Frederiksen} and S.~M. {Kepert}.
\newblock {Dynamical Subgrid-Scale Parameterizations from Direct Numerical
  Simulations}.
\newblock {\em Journal of Atmospheric Sciences}, 63:3006--3019, November 2006.

\bibitem{frederiksen_99}
JS~Frederiksen.
\newblock {Subgrid-scale parameterizations of eddy-topographic force, eddy
  viscosity, and stochastic backscatter for flow over topography}.
\newblock {\em {JOURNAL OF THE ATMOSPHERIC SCIENCES}}, {56}({11}):{1481--1494},
  {JUN 1} {1999}.

\bibitem{Gardiner2009}
Crispin Gardiner.
\newblock {\em {Stochastic Methods}}.
\newblock Springer-Verlag Berlin Heidelberg, 4 edition, 2009.

\bibitem{Haidvogel1999}
Dale~B. Haidvogel and Aike Beckmann.
\newblock {\em {Numerical Ocean Circulation Modeling.}}
\newblock Imperial College Press, London, 1999.

\bibitem{Hasselmann1976}
K.~Hasselmann.
\newblock {Stochastic climate models Part I. Theory}.
\newblock {\em Tellus}, 28(6):473--485, jan 1976.

\bibitem{Honerkamp1994}
J.~Honerkamp.
\newblock {\em Stochastic Dynamical Systems: Concepts, Numerical Methods, Data
  Analysis}.
\newblock Wiley-VCH, 1994.

\bibitem{Imkeller2001}
Peter Imkeller and Jin-Song von Storch, editors.
\newblock {\em {Stochastic Climate Models}}.
\newblock Birkh{\"{a}}user Basel, Basel, 2001.

\bibitem{kha66b}
R.~Z. Khasminsky.
\newblock A limit theorem for the solutions of differential equations with
  random right-hand sides.
\newblock {\em Theory Prob. Applications}, 11:390--406, 1966.

\bibitem{kha66a}
R.~Z. Khasminsky.
\newblock On stochastic processes defined by differential equations with a
  small parameter.
\newblock {\em Theory Prob. Applications}, 11:211--228, 1966.

\bibitem{kitsios_12}
V.~{Kitsios}, J.~S. {Frederiksen}, and M.~J. {Zidikheri}.
\newblock {Subgrid Model with Scaling Laws for Atmospheric Simulations}.
\newblock {\em Journal of Atmospheric Sciences}, 69:1427--1445, April 2012.

\bibitem{kitsios_13}
V.~{Kitsios}, J.~S. {Frederiksen}, and M.~J. {Zidikheri}.
\newblock {Scaling laws for parameterisations of subgrid eddy-eddy interactions
  in simulations of oceanic circulations}.
\newblock {\em Ocean Modelling}, 68:88--105, August 2013.

\bibitem{Kondrashov2005}
Dmitri Kondrashov, S.~Kravtsov, A.~W. Robertson, and M.~Ghil.
\newblock {A hierarchy of data-based ENSO models}.
\newblock {\em Journal of Climate}, 18(21):4425--4444, 2005.

\bibitem{Kraichnan_59}
Robert~H. Kraichnan.
\newblock The structure of isotropic turbulence at very high reynolds numbers.
\newblock {\em Journal of Fluid Mechanics}, 5(4):497--543, 1959.

\bibitem{Kravtsov2005}
S.~Kravtsov, D.~Kondrashov, and M.~Ghil.
\newblock {Multilevel Regression Modeling of Nonlinear Processes: Derivation
  and Applications to Climatic Variability}.
\newblock {\em Journal of Climate}, 18(21):4404--4424, nov 2005.

\bibitem{kur73}
T.~G. Kurtz.
\newblock A limit theorem for perturbed operator semigroups with applications
  to random evolution.
\newblock {\em J. Funct. Anal.}, 12:55--67, 1973.

\bibitem{Lutsko2015}
Nicholas~J. Lutsko, Isaac~M. Held, and Pablo Zurita-Gotor.
\newblock {Applying the Fluctuation–Dissipation Theorem to a Two-Layer Model
  of Quasigeostrophic Turbulence}.
\newblock {\em Journal of the Atmospheric Sciences}, 72(8):3161--3177, 2015.

\bibitem{Majda2002}
A.~Majda, I.~Timofeyev, and E.~Vanden-Eijnden.
\newblock {A priori tests of a stochastic mode reduction strategy}.
\newblock {\em Physica D: Nonlinear Phenomena}, 170(3-4):206--252, 2002.

\bibitem{Majda2009}
Andrew~J Majda, Christian Franzke, and Daan Crommelin.
\newblock {Normal forms for reduced stochastic climate models.}
\newblock {\em Proceedings of the National Academy of Sciences of the United
  States of America}, 106(10):3649--53, 2009.

\bibitem{Majda2001}
Andrew~J. Majda, Ilya Timofeyev, and Eric~Vanden Eijnden.
\newblock {A mathematical framework for stochastic climate models}.
\newblock {\em Communications on Pure and Applied Mathematics}, 54(8):891--974,
  2001.

\bibitem{Majda2003a}
Andrew~J. Majda, Ilya Timofeyev, and Eric Vanden-Eijnden.
\newblock {Systematic Strategies for Stochastic Mode Reduction in Climate}.
\newblock {\em Journal of the Atmospheric Sciences}, 60(14):1705--1722, 2003.

\bibitem{Majewski2002}
Detlev Majewski, D{\"{o}}rte Liermann, Peter Prohl, Bodo Ritter, Michael
  Buchhold, Thomas Hanisch, Gerhard Paul, Werner Wergen, and John Baumgardner.
\newblock {The Operational Global Icosahedral–Hexagonal Gridpoint Model GME:
  Description and High-Resolution Tests}.
\newblock {\em Monthly Weather Review}, 130(2):319--338, 2002.

\bibitem{Monahan2011}
Adam~H. Monahan and Joel Culina.
\newblock {Stochastic averaging of idealized climate models}.
\newblock {\em Journal of Climate}, 24(12):3068--3088, 2011.

\bibitem{Newman2003}
Matthew Newman, Prashant~D. Sardeshmukh, Christopher~R. Winkler, and Jeffrey~S.
  Whitaker.
\newblock {A Study of Subseasonal Predictability}.
\newblock {\em Monthly Weather Review}, 131(8):1715--1732, 2003.

\bibitem{Palmer2001}
T.~N. Palmer.
\newblock {A nonlinear dynamical perspective on model error: A proposal for
  non-local stochastic-dynamic parametrization in weather and climate
  prediction models}.
\newblock {\em Quarterly Journal of the Royal Meteorological Society},
  127(572):279--304, 2001.

\bibitem{palmer2009}
T~N Palmer, R~Buizza, F~Doblas-Reyes, T~Jung, M~Leutbecher, G~J Shutts,
  M~Steinheimer, and A~Weisheimer.
\newblock {Stochastic Parametrization and Model Uncertainty}.
\newblock {\em ECMWF Technical Memoranda}, 598:1--42, 2009.

\bibitem{pap76}
G.~Papanicolaou.
\newblock Some probabilistic problems and methods in singular perturbations.
\newblock {\em Rocky Mountain J. Math}, 6:653--673, 1976.

\bibitem{Pavliotis2008}
Grigorios~A. Pavliotis and Andrew~M. Stuart.
\newblock {\em {Multiscale Methods}}, volume~53 of {\em Texts Applied in
  Mathematics}.
\newblock Springer New York, New York, NY, 2008.

\bibitem{Pegion2011}
Kathy Pegion and Prashant~D. Sardeshmukh.
\newblock {Prospects for Improving Subseasonal Predictions}.
\newblock {\em Monthly Weather Review}, 139(11):3648--3666, 2011.

\bibitem{petoukhov_00}
V.~Petoukhov, A.~Ganopolski, V.~Brovkin, M.~Claussen, A.~Eliseev, C.~Kubatzki,
  and S.~Rahmstorf.
\newblock {CLIMBER-2: A Climate System Model of Intermediate Complexity. Part
  I: Model Description and Performance for Present Climate}.
\newblock {\em Climate Dynamics}, 16:1--17, 2000.

\bibitem{pieroth}
Martin Pieroth, Stamen~I. Dolaptchiev, Matthias Zacharuk, Andrey Gritsun, and
  Ulrich Achatz.
\newblock {Climate-Dependence in Empirical Parameters of Subgrid-Scale
  Parameterizations using the Fluctuation-Dissipation Theorem}.
\newblock {\em Journal of the Atmospheric Sciences}, 2018.
\newblock submitted.

\bibitem{Pope2000}
Stephen~B. Pope.
\newblock {\em {Turbulent Flows}}.
\newblock Cambridge University Press, Cambridge, 2000.

\bibitem{Ripodas2009}
P.~R{\'{i}}podas, A.~Gassmann, J.~F{\"{o}}rstner, D.~Majewski, M.~Giorgetta,
  P.~Korn, L.~Kornblueh, H.~Wan, G.~Z{\"{a}}ngl, L.~Bonaventura, and T.~Heinze.
\newblock {Icosahedral Shallow Water Model (ICOSWM): results of shallow water
  test cases and sensitivity to model parameters}.
\newblock {\em Geoscientific Model Development Discussions}, 2(1):581--638,
  2009.

\bibitem{Satoh2008}
M.~Satoh, T.~Matsuno, H.~Tomita, H.~Miura, T.~Nasuno, and S.~Iga.
\newblock {Nonhydrostatic icosahedral atmospheric model (NICAM) for global
  cloud resolving simulations}.
\newblock {\em Journal of Computational Physics}, 227(7):3486--3514, 2008.

\bibitem{Shutts2005}
Glenn Shutts.
\newblock {A kinetic energy backscatter algorithm for use in ensemble
  prediction systems}.
\newblock {\em Quarterly Journal of the Royal Meteorological Society},
  131(612):3079--3102, 2005.

\bibitem{wrf}
W.C. Skamarock and Coauthors.
\newblock {A Description of the Advanced Research WRF Version 3}.
\newblock Technical Report NCAR/TN-475+STR, NCAR, 2008.

\bibitem{Stull1988}
Roland~B. Stull, editor.
\newblock {\em {An Introduction to Boundary Layer Meteorology}}.
\newblock Springer Netherlands, Dordrecht, 1988.

\bibitem{Vallis2006}
Geoffrey~K Vallis.
\newblock {\em {Atmospheric and Oceanic Fluid Dynamics: Fundamentals and
  Large-Scale Circu- lation}}.
\newblock 2006.

\bibitem{Verkley2011}
W.~T~M Verkley.
\newblock {A maximum entropy approach to the problem of parametrization}.
\newblock {\em Quarterly Journal of the Royal Meteorological Society},
  137(660):1872--1886, 2011.

\bibitem{Verkley2016}
W.~T~M Verkley, P.~C. Kalverla, and C.~A. Severijns.
\newblock {A maximum entropy approach to the parametrization of subgrid
  processes in two-dimensional flow}.
\newblock {\em Quarterly Journal of the Royal Meteorological Society},
  142(699):2273--2283, 2016.

\bibitem{Verkley2014}
W.~T~M Verkley and C.~A. Severijns.
\newblock {The maximum entropy principle applied to a dynamical system proposed
  by Lorenz}.
\newblock {\em European Physical Journal B}, 87(1), 2014.

\bibitem{Vissio2017}
Gabriele Vissio and Valerio Lucarini.
\newblock {A proof of concept for scale-adaptive parameterizations: the case of
  the Lorenz '96 model}.
\newblock {\em Quarterly Journal of the Royal Meteorological Society}, 2017.

\bibitem{Weaver2001}
Andrew~J. Weaver, Michael Eby, Edward~C. Wiebe, Cecilia~M. Bitz, Phil~B. Duffy,
  Tracy~L. Ewen, Augustus~F. Fanning, Marika~M. Holland, Amy MacFadyen,
  H.~Damon Matthews, Katrin~J. Meissner, Oleg Saenko, Andreas Schmittner,
  Huaxiao Wang, and Masakazu Yoshimori.
\newblock {The UVic earth system climate model: Model description, climatology,
  and applications to past, present and future climates}.
\newblock {\em Atmosphere-Ocean}, 39(4):361--428, 2001.

\bibitem{Winkler2001}
Christopher~R. Winkler, Matthew Newman, and Prashant~D. Sardeshmukh.
\newblock {A linear model of wintertime low-frequency variability. Part I:
  Formulation and forecast skill}.
\newblock {\em Journal of Climate}, 14(24):4474--4494, 2001.

\bibitem{wouters2016}
Jeroen Wouters, Stamen {Iankov Dolaptchiev}, Valerio Lucarini, and Ulrich
  Achatz.
\newblock {Parameterization of stochastic multiscale triads}.
\newblock {\em Nonlinear Processes in Geophysics}, 23(6):435--445, 2016.

\bibitem{wouters2012}
Jeroen Wouters and Valerio Lucarini.
\newblock {Disentangling multi-level systems: averaging, correlations and
  memory}.
\newblock {\em Journal of Statistical Mechanics: Theory and Experiment},
  2012(03):P03003, 2012.

\bibitem{wouters2013}
Jeroen Wouters and Valerio Lucarini.
\newblock {Multi-level Dynamical Systems: Connecting the Ruelle Response Theory
  and the Mori-Zwanzig Approach}.
\newblock {\em Journal of Statistical Physics}, 151(5):850--860, 2013.

\end{thebibliography}


\begin{table}[t!]
  \caption{First column in both tables shows the spatially averaged second ($k=2$) and fourth ($k=4$) order centered statistical moments values of DNS. The other columns contain the relative errors of different low-resolution simulations. Upper table for \(H\) and lower table for \(HU\).}\label{t1}
\begin{center}
\begin{tabular}{l | cccccccccccc}
 $H$ & DNS & LRM & BRT & LRM-OU & BRT-OU & BRT-SMR & BRT-SMR-0.6 & BRT-SMR-0.0\\
\hline
 \(k=2\) & 2.846 \((\text{km})^2\) & 0.138 & -0.167 & -0.589 & -0.125 & 0.195 & 0.029 & -0.049 \\ 
 \(k=4\) & 23.56 \((\text{km})^4\) & 0.542 & -0.290 & -0.821 & -0.207 & 0.455 & 0.080 & -0.079    
\end{tabular}
\end{center}
\begin{center}
\begin{tabular}{l | cccccccccccc}
 $HU$ & DNS & LRM & BRT & LRM-OU & BRT-OU & BRT-SMR & BRT-SMR-0.6 & BRT-SMR-0.0\\
\hline
 \(k=2\) & 2082 \((10^3 (\text{km})^2/\text{d})^2\)& 0.255 & -0.159 & -0.597 & -0.141 & 0.099 & -0.052 & -0.124 \\ 
 \(k=4\) & 1.277E+07 \((10^3 (\text{km})^2/\text{d})^4\) & 1.119 & -0.274 & -0.834 & -0.251 & 0.185 & -0.119 & -0.247    
\end{tabular}
\end{center}
\end{table}

\begin{table}[t!]
\caption{Summary of the different models used}\label{t2}
\begin{center}
\begin{tabular}{ll}
\hline
 DNS & Direct Numerical Simulation with spatial resolution of \(N\) cells\\
     & \(\dot{x}_i =  \varrho^x_i +  a^x_i(\textbf{x}) +  b^{xz}_i(\textbf{x},\textbf{z})\); \(\dot{z}_i = b^z_i(\textbf{x},\textbf{z}) + c^z_i(\textbf{z})\) \\ 
\hline
 BRT & Bare-Truncation Model with spatial resolution of \(N_c=N/n\) \\
     &\(\dot{x}_i =  \varrho^x_i +  a^x_i(\textbf{x})\) \\  
\hline
 LRM & Low-Resolution Model \\ 
     & DNS with spatial resolution of \(Nc=N/n\).\\
\hline
 OU-DNS & DNS with \(\frac{1}{h}\)-approximation 
          and SGS self-interactions replaced by an OU process \\
& \(\dot{x}_i = \varrho^x_i + a^x_i (\textbf{x}) +  b^{x}_i(\textbf{x},\textbf{y})\), \(\dot{y}_i = b^y_i(\textbf{x},\textbf{y}) + \Lambda_{ij} y_j + \Sigma_i \dot{W}_i \). \\
\hline
 BRT-OU & BRT with empirical Ornstein-Uhlenbeck parameterization \\
     & \(dx_i = \left(\varrho^x_i + a^x_i (\textbf{x}) + \tilde{\Gamma}_{ij} \hat x^I_j \right )dt + \tilde{\sigma_i} dW_i\). \\
\hline
 LRM-OU & LRM with empirical Ornstein-Uhlenbeck parameterization \\
\hline
 BRT-SMR & BRT with Stochastic Mode Reduction parameterization \\
         & \(dx_i = \left[ \varrho^x_i + a^x_i(\textbf{x}) + \beta_i(\textbf{x}) \right] dt + d\xi_{i}(\textbf{x})\)\\
\hline
 BRT-SMR-0.6 & BRT-SMR with stochastic forcing reduced to \(60 \%\) \\
&  \(dx_i = \left[ \varrho^x_i + a^x_i(\textbf{x}) + \beta_i(\textbf{x}) \right] dt + 0.6 d\xi_{i}(\textbf{x})\).\\
\hline
BRT-SMR-0.0 & BRT-SMR without stochastic forcing \\
&  \(dx_i = \left[ \varrho^x_i + a^x_i(\textbf{x}) + \beta_i(\textbf{x}) \right] dt\)\\
\hline
\end{tabular}
\end{center}
\end{table}


\end{document}